\documentclass{aac}

\usepackage[numbers]{natbib}
\usepackage{rotating}
\usepackage{graphicx}

\usepackage{xcolor}

\begin{document}

\title{Nonlinear Inverse Compton Scattering from a Laser Wakefield Accelerator and Plasma Mirror}

\author[1]{A. Hannasch}
\author[1,2]{M. LaBerge}
\author[1]{R. Zgadzaj}
\author[2]{J. P. Couperus Cabada\u{g}}
\author[2]{A. Laso Garcia}
\author[2]{T. Kurz}
\author[2]{T. Cowan}
\author[2]{U. Schramm}
\author[2]{A. Irman}
\author[1]{M.C. Downer\corref{cor1}}

\affil[1]{The University of Texas at Austin, Department of Physics, Austin, Texas 78712-1081, USA.}
\affil[2]{The Helmholtz-Zentrum Dresden-Rossendorf, Institute for Radiation Physics, 01328 Dresden, Germany}
\corresp[cor1]{downer@physics.utexas.edu}

\maketitle

\begin{abstract}
We generate inverse Compton scattered X-rays in both linear and nonlinear regimes with a 250 MeV laser wakefield electron accelerator and plasma mirror by retro-reflecting the unused drive laser light to scatter from the accelerated electrons. We characterize the X-rays using a CsI(Tl) voxelated scintillator that measures their total energy and divergence as a function of plasma mirror distance from the accelerator exit. At each plasma mirror position, these X-ray properties are correlated with the measured fluence and inferred intensity of the laser pulse after driving the accelerator to determine the laser strength parameter $a_0$. The results show that ICS X-rays are generated at $a_0$ ranging from $0.3 \pm 0.1$ to $1.65 \pm 0.25$, and exceed the strength of co-propagating bremsstrahlung and betatron X-rays at least ten-fold throughout this range of $a_0$. 
\end{abstract}

\section{INTRODUCTION}
The discovery and development of X-ray sources has spurred progress over the last 100 years in material science \cite{Jaeschke2016SynchrotronLasers}, geology \cite{Ketcham2001AcquisitionGeosciences}, industry \cite{Hanke2008X-rayCharacterization}, diagnostic medicine \cite{Lewis1997MedicalX-rays,Suortti2003MedicalRadiation},  high-Z radiography \cite{Chen2007Dual-energyDetection} and probing warm dense matter (WDM) \cite{Falk2014CombinedShock-and-release}. Many such X-ray sources require meter- to kilometer-scale radio frequency (RF) accelerators to generate synchrotron or bremsstrahlung radiation. Laser plasma accelerators (LPAs) \cite{Tajima1979LaserAccelerator,Esarey2009PhysicsAccelerators} offer a compact alternative, capable of accelerating electrons up to MeV energies over millimeters or GeV energies over centimeters \cite{Wang2013Quasi-monoenergeticGeV, Gonsalves2019PetawattWaveguide} with narrow bandwidth (1-15\%) and 100s of pC of charge \cite{Couperus2017DemonstrationAccelerator,Gotzfried2020PhysicsWakefields}. When the accelerated electrons scatter from a counter-propagating laser pulse, they transfer energy to the photons via inverse Compton scattering (ICS) \cite{Corde2013FemtosecondAccelerators}. The laser strength parameter $a_0 \approx 0.85\lambda[\mu \mbox{m}]\sqrt{I[10^{18}\,\mbox{W/cm}^2]}$ plays a significant role in X-ray generation via ICS and defines the scattering regime. For $a_0 \ll 1$, the emitted radiation is narrowband and peaks at $\approx 4 \gamma_e^2 \hbar \omega_L$ where $\gamma_e$ is the electron Lorentz factor and $\hbar \omega_L$ is the energy of the scattering laser photons. As $a_0$ approaches 1 the scattering becomes relativistic as the laser imparts significant momentum to the electrons, resulting in high flux and broadband X-rays \cite{Corde2013FemtosecondAccelerators,Kramer2018MakingApplications,Khrennikov2015TunableRegime}. Such tabletop LPA-based ICS sources, or ``all-optical undulators", can generate hard X-ray pulses comparable in peak brilliance to those from large-scale synchrotrons \cite{Corde2013FemtosecondAccelerators}.

There are two methods for generating ICS X-ray beams from a compact LPA. The first method uses two separate laser pulses; the first drives the accelerator while the second, split from the drive pulse before acceleration, scatters nearly head-on with the accelerated electron bunch. 
The secondary pulse can reach $a_0 > 4$ when focused tightly with low f/\# optics to the collision point \cite{Yan2017High-orderScattering, Cole2018ExperimentalBeam, Sarri2014UltrahighScattering}, but is difficult to align and borrows energy from the LPA drive pulse. The second method for generating ICS uses a single laser pulse that first drives the LPA and then reflects from a plasma mirror (PM) to scatter from the trailing relativistic electrons \cite{TaPhuoc2012All-opticalSource,Tsai2015CompactMirror}. Because the LPA drive pulse leads the electrons by no more than tens of microns the reflected pulse is automatically overlapped with the electrons. This "self-aligning" feature makes ICS highly reproducible, ideal for use as an electron beam diagnostic \cite{Kramer2018MakingApplications} or an experimental probe without requiring an additional laser pulse or reducing the drive laser energy. It is therefore the method of choice for smaller LPA laboratories with limited available laser power and funds. However, the properties of ICS X-rays generated in this way have not been systematically characterized. In particular, the ability of this method to reach the important relativistic ICS regime ($a_0 > 1$) has not previously been demonstrated.  

\begin{figure}[ht!]
\centerline{\includegraphics[width=.85\textwidth]{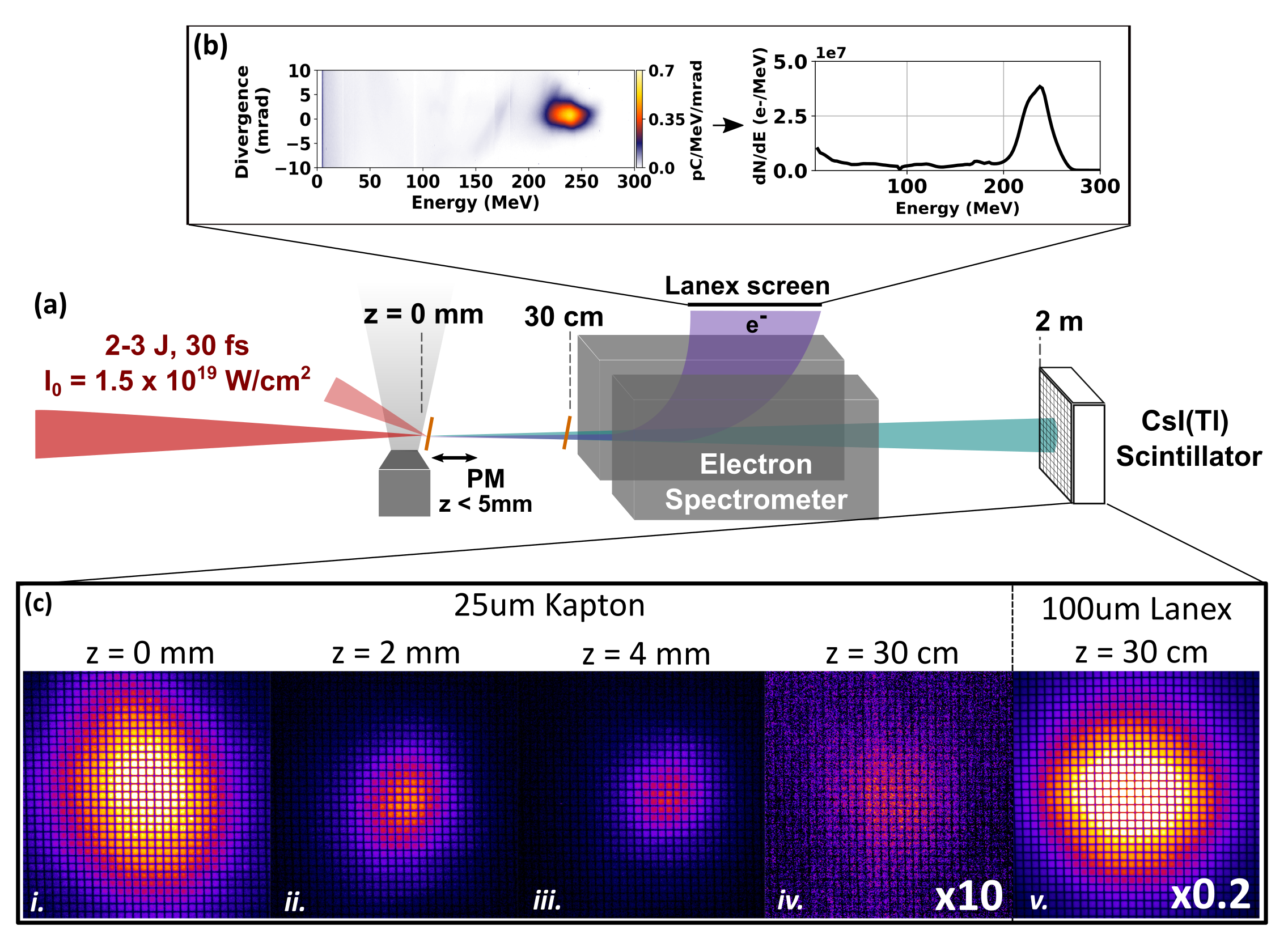}}
\caption{\label{fig:setup} LPA X-ray characterization overview.  (a) Schematic set up showing (left to right) incident laser pulse, gas jet, tilted plasma mirror (PM) positioned at $0 < z < 0.5$\,cm from gas jet exit for generating ICS X-rays, converter at $z \approx 30$\,cm for generating bremsstrahlung, 1 T magnetic electron spectrometer, and 3 cm thick CsI(Tl) profiling scintillator outside of vacuum. (b) Representative single-shot electron spectrum. Left:  raw data from luminescent Lanex screen.  Right:  electron energy distribution averaged over emission angle. (c) i.-iii. ICS profiles for a 25-$\mu$m-thick Kapton PM at varying positions after the accelerator and iv.-v. bremsstrahlung profiles generated with 25-µm-thick Kapton (iv) or 100-µm-thick Lanex (v) placed 30 cm after the accelerator. Multiplicative factors indicate what each profile signal is multiplied by in order to directly compare with the ICS profiles.}
\end{figure}

Here, we characterize an ICS X-ray source based on a 250 MeV laser wakefield electron accelerator and a plasma mirror in both linear and nonlinear scattering regimes. The laser fluence is measured after driving the LPA and the radiative properties of the ICS source are characterized with a compact scintillating detector. We demonstrate that the scattering regime can be easily tuned by changing the placement of the PM relative to the gas jet.

\subsection{PROCEDURE}

Experiments were performed on the  DRACO Ti:Sa chirped pulse amplification laser system at Helmholtz-Zentrum Dresden-Rossendorf (HZDR) \cite{Irman2018ImprovedInjection} which can deliver up to $\sim$ 4 J, 30 fs pulses with a spot size of 20 $\mu$m (FWHM) to a 3 mm long He gas jet with 1 \% Nitrogen doping [see Fig.\,\ref{fig:setup}(a)]. The accelerator operates in a self-truncated ionization injection regime \cite{Couperus2017DemonstrationAccelerator, Mirzaie2015DemonstrationBeams} and the typical parameters for the experiments presented here are 2 J on target and electron density of $4 \times 10^{18}$ $\mathrm{cm^{-3}}$. A magnetic spectrometer placed 30 cm downstream from the gas jet sweeps accelerated electrons off-axis vertically and measures their spectrum and horizontal divergence \cite{Schramm2017FirstDresden,Kurz2018CalibrationDetermination} [see Fig.\,\ref{fig:setup}(b)]. The profile and fluence of the laser pulse were characterized by picking off the drive laser after the magnet with no PM in place and imaging the profile to a camera outside the target chamber under acceleration conditions with and without gas. The total laser energy on target was calibrated with a power meter and the beam size was calibrated by imaging a 20 $\mu$m tungsten wire placed at the exit of the gas jet. The laser fluence is characterized at selected positions downstream of the accelerator by integrating the camera signal within a calibrated radius of 4 $\mu$m at the maximum. This area is chosen to best match the interaction area of the electrons with the reflected laser. The ICS profile and total radiated energy are diagnosed at an electron density of $4 \times 10^{18}~\mathrm{cm^{-3}}$ by a $7 \times 7 \times 3$ cm CsI(Tl) profiling scintillator placed on-axis 2 m after the gas jet and outside of vacuum. Scintillating crystal detectors are well-suited for detecting X-rays between 0.1 and several MeV photon energy because of their high Z, high photon output and relatively flat spectral response in this range. They measure changes in total radiated energy accurately. To determine the range of $a_0$ over which we can generate ICS photons with a PM, we varied the position of the PM between $z=-0.2$ mm (inside the accelerator) and $z=4$ mm downstream and imaged the total scintillator signal with a CCD camera. Figure \ref{fig:setup}(c) shows the ICS profile from a 25 $\mu$m Kapton PM placed 0 (i.),  2 (ii.), and  4 (iii.) mm after the accelerator. The profiles generated by a target placed at the entrance of the magnet (z = 30 cm) are dominated by bremsstrahlung and are multiplied by the factor shown to compare to the ICS shots on the same scale. Bremsstrahlung from a  25 $\mu$m Kapton film (iv.) is $\sim 10\times$ less bright than the ICS signal, while bremsstrahlung from a 100 $\mu$m Lanex screen (GdS:Tb, v.) is $> 5\times$ brighter. Betatron radiation can also deposit energy in the scintillator, however no observed signal is recorded by the scintillating camera when the LPA is on but no material is placed in the accelerated electron beam path.

\section{RESULTS}
\subsection{Laser fluence}

\begin{figure}[ht!]
\centerline{\includegraphics[width=.85\textwidth]{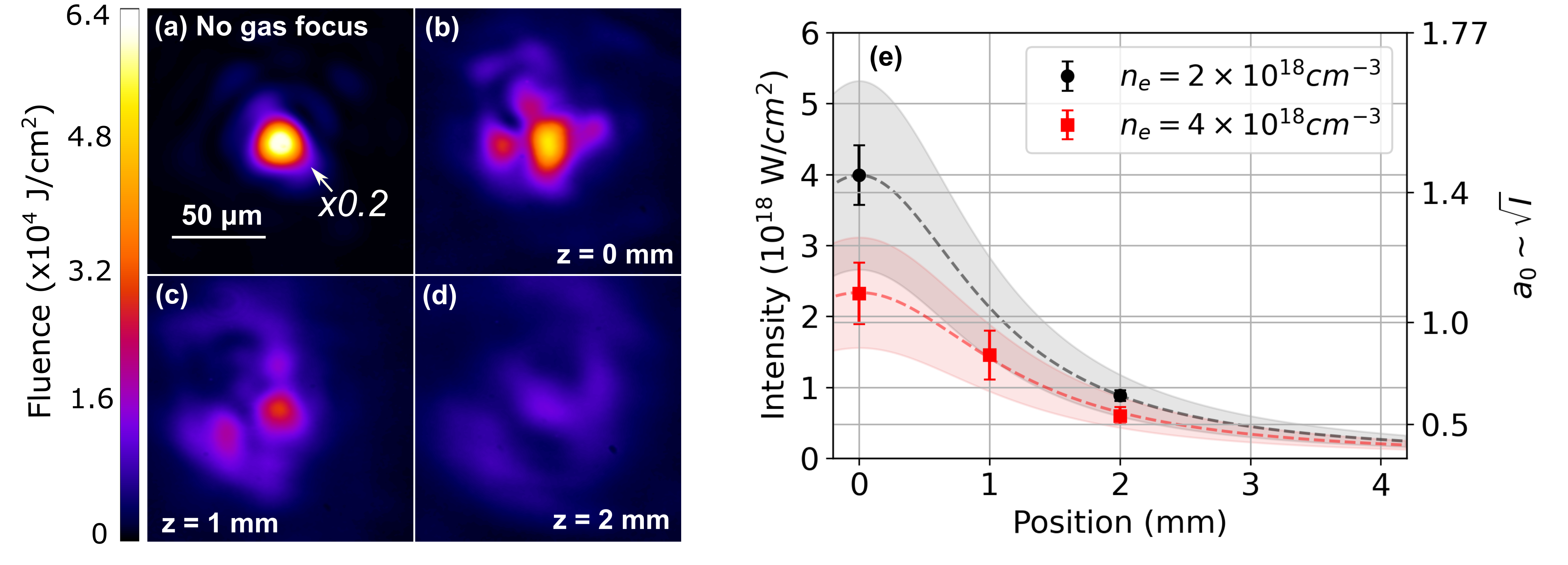}}
\caption{\label{fig:Laser intensity} The imaged laser focus in vacuum is presented in (a) and is 20 $\mu$m FWHM and $2\times 10^{19} ~\mbox{W/cm}^2$. Panels (b)-(d) show characteristic examples of the imaged laser profile after acceleration for a density $n_e = 4\times 10^{18}~\mathrm{cm^{-3}}$ at (b) $z=0$, (c) $z=1$ and (d) $z=2$ mm after the gas jet exit. (e) The laser intensity is inferred from the measured fluence at multiple locations after the accelerator exit for pulse duration of 20 fs and a density of $n_e = 2\times 10^{18}~\mathrm{cm^{-3}}$ (black circles) and $n_e = 4\times 10^{18}~\mathrm{cm^{-3}}$ (black circles). }
\end{figure}

Figure \ref{fig:Laser intensity} shows the (a) vacuum focus laser profile with a FWHM beam size of 20 $\mu$m, the laser profile after propagating through 3 mm of He at a density $n_e = 4\times 10^{18}~\mathrm{cm^{-3}}$ at (b) $z=0$, (c) $z=1$ mm ,and (d) $z=2$ mm downstream of the gas jet exit. Before acceleration, the laser has a FWHM pulse duration of 30 fs, peak intensity of $2\times 10^{19}~ \mbox{W/cm}^2$ and $a_0 \approx 3$. During acceleration, the laser loses energy to the plasma and the spatial profile modulates, however, the pulse duration is expected to decrease as the front edge of the laser etches back \cite{Decker1996ThePlasmas,Zhu2012StudiesSimulations}. Simulations predict that under the acceleration conditions presented here the laser pulse can decrease to as low as $\sim10-15$ fs. Figure \ref{fig:Laser intensity}(e) shows the calculated laser intensity based on the measured fluence for a pulse duration of 20 fs and density of $n_e=2\times 10^{18}~\mathrm{cm^{-3}}$ (black circles) and $4\times 10^{18}~\mathrm{cm^{-3}}$ (red squares). The error bars represent the statistical error from the standard deviation over $\sim 10$ shots and the shaded regions give the range from the lower bound intensity which assumes a pulse duration of 30 fs, and an upper bound intensity that assumes a pulse duration of 15 fs. The calculated intensity after acceleration (data points) is fit to $(z^2+z_r^2)^{-1}$ which assumes a local Gaussian propagation of Rayleigh length $z_r$ once the laser leaves the gas jet. These fits are shown by the dotted lines for both densities. 

\subsection{Total Radiated Energy}

\begin{figure}[ht!]
\centerline{\includegraphics[width=.5\textwidth]{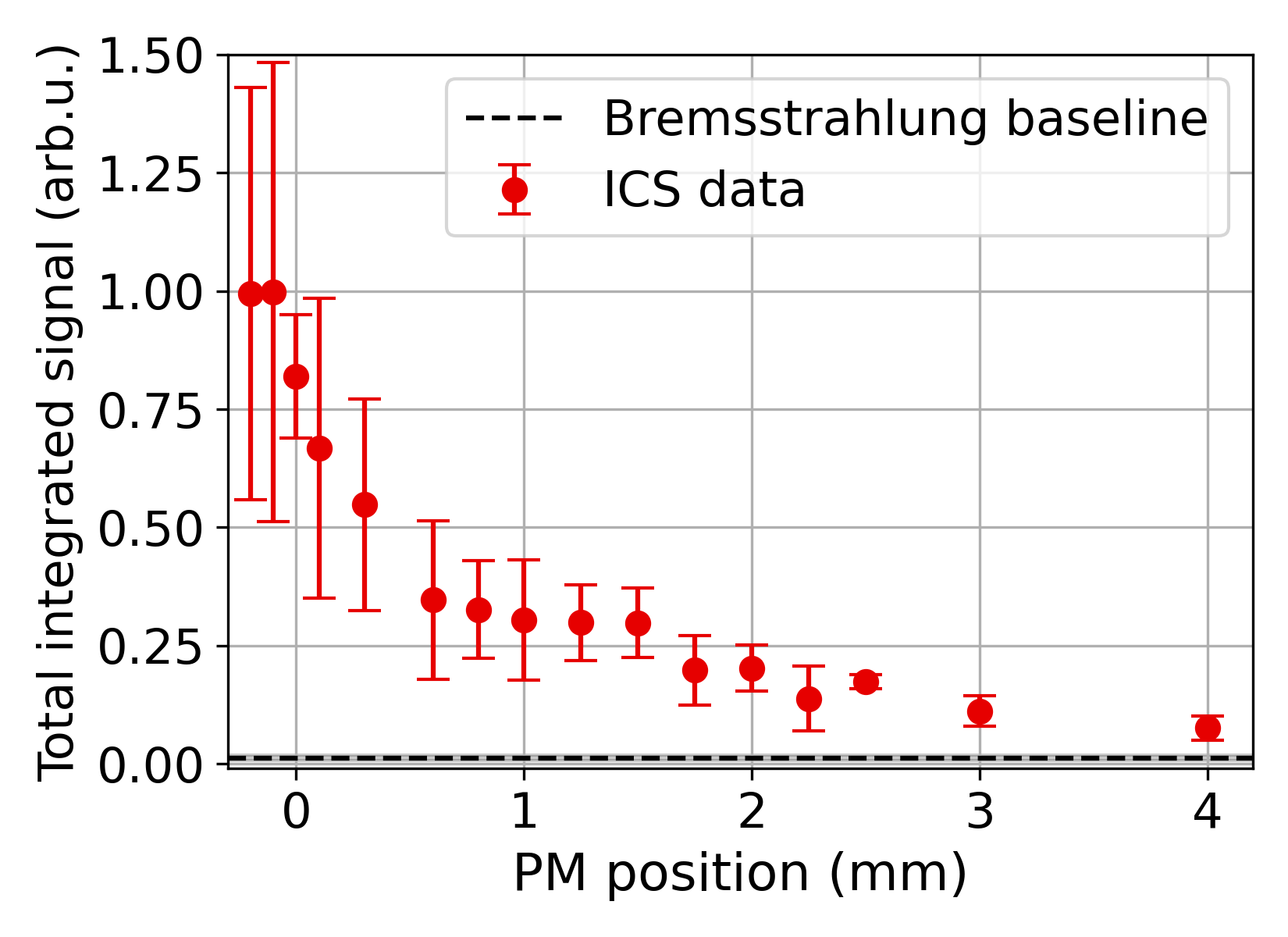}}
\caption{\label{fig:ICS intensity}Total integrated scintillator signal for each position of the plasma mirror averaged over $\sim$5 shots for a density $n_e=4\times 10^{18}~\mathrm{cm^{-3}}$. The bremsstrahlung baseline is determined by the total integrated scintillator signal when the PM is placed at z = 30 cm where the laser is not intense enough to generate ICS radiation.}
\end{figure}

Figure \ref{fig:ICS intensity} shows the background subtracted total integrated scintillator signal averaged over $\sim5$ shots (red circles) for each position of the PM relative to the exit of the gas jet at $n_e = 4\times 10^{18}~\mathrm{cm^{-3}}$. The data is scaled to the maximum value and the black dashed line provides the normalized scintillator signal when the PM was moved 30 cm downstream, indicating a baseline for the bremsstrahlung signal generated through the 25 $\mu$m-thick Kapton film. The weakest signal generated via ICS with the PM at z = 4 mm has $\sim 6 \times$ more energy than this baseline, indicating that the measurements are dominated by the ICS contribution. Furthermore we observe a $\sim 13 \times$ increase in the signal when the PM is placed within the jet exit versus 4 mm downstream offering a simple but robust tunability of the scattering regime.

\subsection{Divergence}

\begin{figure}[ht!]
\centerline{\includegraphics[width=.88\textwidth]{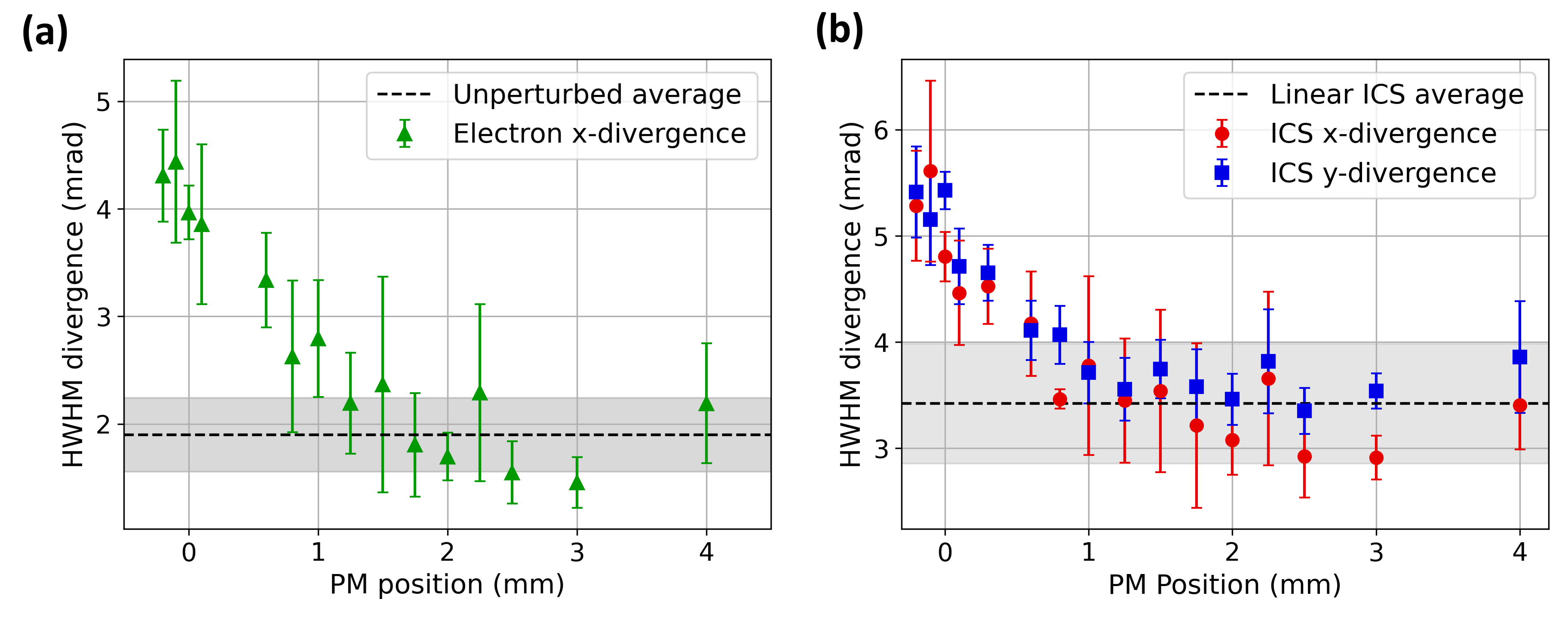}}
\caption{\label{fig:Divergence plots}Divergence of the (a) electrons in the horizontal plane of the electron spectrometer as a function of plasma mirror position after the accelerator exit at $n_e=4\times 10^{18}~\mathrm{cm^{-3}}$. The dotted black line represents the unperturbed average electron divergence with no plasma mirror in place and the shaded region corresponds to the standard deviation. (b) The divergence of the ICS profiles for the corresponding shots in the vertical dimension (blue squares) and horizontal dimension (red circles). The dotted line represents the average divergence for PM placed $z>1.5$ mm which most closely corresponds to a linear scattering regime.}
\end{figure}

Figure \ref{fig:Divergence plots}(a) shows the HWHM divergence of the electrons for different plasma mirror positions (green triangles) averaged over $\sim$5 shots. The dashed line represents the average HWHM divergence of electrons without the PM in place of $1.9 \pm 0.3$ mrad. The shaded bar shows the standard deviation of the unperturbed electrons to indicate the shot to shot fluctuations during the experiment. We observe an increase of $\sim 2 \times$ in the electron divergence when the PM is close to the jet ($<$ 1 mm) resulting from strong magnetic fields within the plasma mirror at relativistic laser intensities \cite{Raj2020ProbingInteractions,Schumaker2013UltrafastInteractions}. 
The perturbation from the electrons passing through the PM bulk occurs \emph{after} the reflected laser scatters from the accelerated electrons, thus we can assume that the electron divergence remains constant at $\sim 1.9 \pm 0.3$ mrad during the ICS interaction.

Figure \ref{fig:Divergence plots}(b) shows the HWHM divergence of the ICS X-rays as a function of plasma mirror position where we observe divergence growth in the X-rays as the PM is moved $\leq 1$ mm. The linear ICS divergence can be found by averaging the divergence of the X-ray beam when the PM is placed at $z \geq 1.5$ mm downstream. The result is $3.4 \pm 0.6$ mrad HWHM, indicated by the dashed line and shaded region in figure \ref{fig:Divergence plots}(b). The ICS divergence is up to $\sim 60\%$ larger than this baseline when the PM is place $< 0.5$ mm after the gas jet and, assuming the electron divergence remains constant, this X-ray divergence growth is caused by nonlinear ICS scattering effects. Expectations of the ICS divergence dependence on electron divergence will be discussed in more detail in the Discussion.

\section{DISCUSSION}



Understanding the strength of the laser during the scattering process is important for predicting the scattering regime and diagnosing the emitted radiation. For applications that require a narrowband source of X-rays, the laser strength parameter $a_0$ must remain small ($a_0 \ll 1$) so that the interaction with the accelerated electrons is linear. If a higher flux source is needed and a narrowband source is not, then a nonlinear interaction is required ($a_0 \geq 1$). Estimates can be made based on the measured laser fluence prior to PM reflection and the resulting relative integrated signal observed with the profiling scintillator. The expected energy radiated from a single relativistic electron is proportional to $\gamma_e^2$, and when integrated over the total electron spectrum $E_{tot} \propto \int{\gamma^2 \frac{dN_e}{d\gamma_e} d\gamma_e = N_b\left<\gamma_e^2\right>}$. Inverse Compton scattered radiation is additionally proportional to $a_0^2\propto I_L$ \cite{Esarey1993NonlinearPlasmas} such that $E_{tot} \propto N_b \left<\gamma_e^2\right> a_0^2$. The integrated scintillator signal, which is roughly proportional to the total radiated energy of the ICS source, can be normalized to the electron bunch parameter $N_e \left< \gamma_e^2 \right>$ which reduces the shot-to-shot variation observed in Fig.\,\ref{fig:ICS intensity} and isolates the effect $a_0$ has on the resulting ICS radiation. 

\begin{figure}[ht!]
\centerline{\includegraphics[width=.95\textwidth]{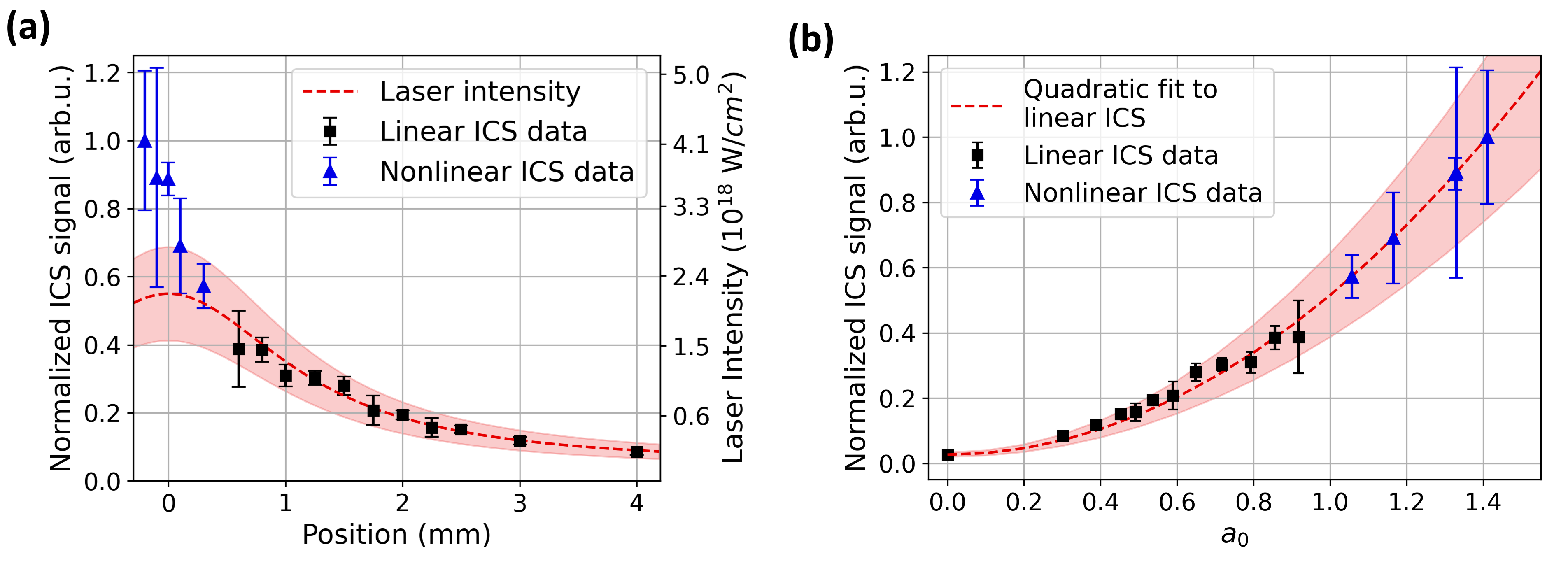}}
\caption{\label{fig:Laser ICS plots}(a) Combined plot of the calculated scattering laser intensity after reflecting from the PM and the resulting total ICS signal as measured from the CsI(Tl) scintillator. The corresponding laser intensity for the linear ICS data (black squares, PM @ $z>0.5$ mm) was converted to $a_0$ and fit to a quadratic model in (b) and to provide a method for interpolating the scattering intensity for the nonlinear ICS data (blue triangles, PM @ $z<0.5$ mm).}
\end{figure}

Figure \ref{fig:Laser ICS plots}(a) shows the \emph{normalized} ICS integrated scintillator signal (data points) on the left axis and the calculated laser intensity (red dashed curve) on the right axis. The laser intensity is calculated from the measured fluence, assuming a 20 fs pulse duration and an 85\% reflectivity efficiency of the PM \cite{Tsai2017Self-aligningFocus}. The shaded region represents a $\pm 25\%$ combined uncertainty in the pulse duration and reflectivity of the PM and the two sets of data are linearly fit to find the corresponding scattering intensity at each PM position. However, this procedure only allows for an estimate of the ICS scattering intensity for PM positions $z>0.5$ mm, referred to as "Linear ICS data" in Fig.\,\ref{fig:Laser ICS plots}(a) (black squares). For PM positions $z<0.5$ mm, the ICS intensity deviates from the laser intensity curve, suggesting that the real scattering intensity and $a_0$ is larger than that estimated by the laser intensity curve alone. This observation may indicate a nonlinear response of the PM to the incident laser intensity. Simulations and experiments have shown that a laser pulse with $a_0 \approx 1$ induces a curvature at the PM surface and focuses the laser pulse with a focal length as small as 25 $\mu$m, resulting in a larger intensity of the reflected pulse compared to the incident pulse \cite{Tsai2017Self-aligningFocus}.

To estimate the value of $a_0$ at PM positions near the gas jet exit (Fig.\,\ref{fig:Laser ICS plots}(a) "Nonlinear ICS data", blue triangles), the linear ICS data can be plotted as a function of $a_0$ and a quadratic fit can be performed, including the normalized bremsstrahlung baseline as the y-intercept. The resulting quadratic function that relates the normalized integrated scintillator signal and the corresponding values of $a_0$ from Fig.\,\ref{fig:Laser ICS plots}(a) enables extrapolation of the nonlinear ICS $a_0$ values. Figure \,\ref{fig:Laser ICS plots}(b) illustrates this procedure where the "Nonlinear ICS data" for PM positions $z < 0.5$ mm (Fig.\,\ref{fig:Laser ICS plots}(b) blue triangles) is extrapolated based on the quadratic fit of the "Linear ICS data" to $a_0$ (Fig.\,\ref{fig:Laser ICS plots}(b) red dashed curve). This extrapolation indicates that the intensity of the scattering laser can reach up to $a_0=1.4 \pm 0.2$. The possible PM focusing effect suggests that the PM method for generating ICS radiation is not wholly limited by the intensity of the laser after driving the acceleration process, and is capable of generating radiation in both the linear ($a_0 \ll 1$) and nonlinear ($a_0 \geq 1$) ICS regime. 

The observed divergence growth when the PM is placed $z < 0.5$ further supports the conclusion that we observe both linear and nonlinear ICS from an LPA and PM. ICS from a single electron has a HWHM divergence $\theta_x \approx 1/\gamma_e$ for $a_0 \ll 1$ and increases to $\theta_x \approx a_0/\gamma_e$ when $a_0 > 1$ \cite{Corde2013FemtosecondAccelerators}. When scattering from an electron bunch with HWHM divergence $\theta_e$, the resulting ICS divergence can be approximated as $\theta_{ICS} = \sqrt{\theta_e^2 + \theta_x^2}$. Assuming the scattering electron divergence remains constant for each PM position, the expected linear ICS divergence based on this equation is $2.9 \pm 0.8$ and agrees with the measured linear ICS divergence to within uncertainty. The observed divergence growth of the ICS when the PM is placed $z<0.5$ mm is a result of nonlinear ICS when the value of $a_0$ approaches and exceeds 1. This nonlinear divergence growth has been observed in ICS experiments using a separate colliding pulse instead of a PM \cite{Yan2017High-orderScattering}. Finally, the unperturbed electron divergence and the linear ICS divergence can be used to estimate an upper bound value of $a_0$ for the nonlinear ICS data of $1.9 \pm 0.2$. 

\section{CONCLUSION}
We have presented scintillator based measurements of the nonlinear characteristics of ICS radiation from a 250 MeV LPA and plasma mirror. The total radiated energy of the linear ICS beam is $\sim 6 \times$ larger than the bremsstrahlung generated from the 25 $\mu$m Kapton plasma mirror when placed 4 mm after the accelerator exit. Moreover, we observe $\sim 13 \times$ more energy radiated when the PM is placed 200 $\mu$m within the gas jet compared to 4 mm downstream indicating a $\sim 3.6$ increase in the scattering parameter $a_0$. Measurements of the laser fluence and inferred intensity show that the laser remains relativistic with $a_0$ of $1.2 \pm 0.2$ after driving the LPA and reduces to $a_0 \sim 0.3 \pm 0.1$ after propagating 4 mm from the gas jet. By assuming a PM reflectivity of 0.85 and pulse duration of 20 fs we are able to fit the integrated scintillator signal from the linear ICS to the corresponding laser intensity, providing a method for extrapolating the value of $a_0$ when the PM is placed $< 0.5$ mm. We find that the scattering laser intensity could reach as high as $a_0$ of $1.4-1.9$ when the PM is placed within the exit of the gas jet. This increase is also more than would be suggested by the measured fluence of the laser pulse and may be caused by an induced curvature of the PM by the intense laser pulse resulting in a focused and higher reflected intensity at the point of interaction with the electrons. These measurements are supported by the increase in the ICS divergence from the linear to nonlinear ICS scattering regime. To the best of our knowledge this is the first observation of both linear and nonlinear ICS from a laser plasma accelerator and plasma mirror.

\section{ACKNOWLEDGMENTS}

U.\,Texas authors acknowledge support from U. S. Department of Energy grant DE-SC0011617, A.H. from the National Science Foundation Graduate Research Fellowship grant No. DGE-1610403 and M. C. D. from the Alexander von Humboldt Foundation.  HZDR authors acknowledge support from the Helmholtz Association under program Matter and Technology, topic Accelerator R \& D. 


\bibliographystyle{aac}%
\bibliography{references}%

\begin{thebibliography}{31}%
\makeatletter
\providecommand \@ifxundefined [1]{%
 \@ifx{#1\undefined}
}%
\providecommand \@ifnum [1]{%
 \ifnum #1\expandafter \@firstoftwo
 \else \expandafter \@secondoftwo
 \fi
}%
\providecommand \@ifx [1]{%
 \ifx #1\expandafter \@firstoftwo
 \else \expandafter \@secondoftwo
 \fi
}%
\providecommand \natexlab [1]{#1}%
\providecommand \enquote  [1]{``#1''}%
\providecommand \bibnamefont  [1]{#1}%
\providecommand \bibfnamefont [1]{#1}%
\providecommand \citenamefont [1]{#1}%
\providecommand \href@noop [0]{\@secondoftwo}%
\providecommand \href [0]{\begingroup \@sanitize@url \@href}%
\providecommand \@href[1]{\@@startlink{#1}\@@href}%
\providecommand \@@href[1]{\endgroup#1\@@endlink}%
\providecommand \@sanitize@url [0]{\catcode `\$12\catcode `\&12\catcode
  `\#12\catcode `\^12\catcode `\_12\catcode `\%12\relax}%
\providecommand \@@startlink[1]{}%
\providecommand \@@endlink[0]{}%
\providecommand \url  [0]{\begingroup\@sanitize@url \@url }%
\providecommand \@url [1]{\endgroup\@href {#1}{\urlprefix }}%
\providecommand \urlprefix  [0]{URL }%
\providecommand \Eprint [0]{\href }%
\providecommand \doibase [0]{http://dx.doi.org/}%
\providecommand \selectlanguage [0]{\@gobble}%
\providecommand \bibinfo  [0]{\@secondoftwo}%
\providecommand \bibfield  [0]{\@secondoftwo}%
\providecommand \translation [1]{[#1]}%
\providecommand \BibitemOpen [0]{}%
\providecommand \bibitemStop [0]{}%
\providecommand \bibitemNoStop [0]{.\EOS\space}%
\providecommand \EOS [0]{\spacefactor3000\relax}%
\providecommand \BibitemShut  [1]{\csname bibitem#1\endcsname}%
\let\auto@bib@innerbib\@empty
\bibitem [{\citenamefont {Jaeschke}\ \emph {et~al.}(2016)\citenamefont
  {Jaeschke}, \citenamefont {Khan}, \citenamefont {Schneider},\ and\
  \citenamefont {Hastings}}]{Jaeschke2016SynchrotronLasers}%
  \BibitemOpen
  \bibfield  {author} {\bibinfo {author} {\bibfnamefont {E.~J.}\ \bibnamefont
  {Jaeschke}}, \bibinfo {author} {\bibfnamefont {S.}~\bibnamefont {Khan}},
  \bibinfo {author} {\bibfnamefont {J.~R.}\ \bibnamefont {Schneider}}, \ and\
  \bibinfo {author} {\bibfnamefont {J.~B.}\ \bibnamefont {Hastings}},\ }\href
  {\doibase 10.1007/978-3-319-14394-1} {\emph {\bibinfo {title} {Springer
  Reference}}}\ (\bibinfo {year} {2016})\BibitemShut {NoStop}%
\bibitem [{\citenamefont {Ketcham}\ and\ \citenamefont
  {Carlson}(2001)}]{Ketcham2001AcquisitionGeosciences}%
  \BibitemOpen
  \bibfield  {author} {\bibinfo {author} {\bibfnamefont {R.~A.}\ \bibnamefont
  {Ketcham}}\ and\ \bibinfo {author} {\bibfnamefont {W.~D.}\ \bibnamefont
  {Carlson}},\ }\href {\doibase 10.1016/S0098-3004(00)00116-3} {\bibfield
  {journal} {\bibinfo  {journal} {Computers and Geosciences}\ }\textbf
  {\bibinfo {volume} {27}},\ \unskip\ \bibinfo {pages} {381--400} (\bibinfo
  {year} {2001})}\BibitemShut {NoStop}%
\bibitem [{\citenamefont {Hanke}, \citenamefont {Fuchs},\ and\ \citenamefont
  {Uhlmann}(2008)}]{Hanke2008X-rayCharacterization}%
  \BibitemOpen
  \bibfield  {author} {\bibinfo {author} {\bibfnamefont {R.}~\bibnamefont
  {Hanke}}, \bibinfo {author} {\bibfnamefont {T.}~\bibnamefont {Fuchs}}, \ and\
  \bibinfo {author} {\bibfnamefont {N.}~\bibnamefont {Uhlmann}},\ }\href
  {\doibase 10.1016/j.nima.2008.03.016} {\bibfield  {journal} {\bibinfo
  {journal} {Nuclear Instruments and Methods in Physics Research, Section A:
  Accelerators, Spectrometers, Detectors and Associated Equipment}\ }\textbf
  {\bibinfo {volume} {591}},\ \unskip\ \bibinfo {pages} {14--18} (\bibinfo
  {year} {2008})}\BibitemShut {NoStop}%
\bibitem [{\citenamefont {Lewis}(1997)}]{Lewis1997MedicalX-rays}%
  \BibitemOpen
  \bibfield  {author} {\bibinfo {author} {\bibfnamefont {B.}~\bibnamefont
  {Lewis}},\ }\href {\doibase 10.1080/08940886.2011.567156} {\bibfield
  {journal} {\bibinfo  {journal} {Physics in Medicine and Biology}\ }\textbf
  {\bibinfo {volume} {42}},\ \unskip\ \bibinfo {pages} {1213--1243} (\bibinfo
  {year} {1997})}\BibitemShut {NoStop}%
\bibitem [{\citenamefont {Suortti}\ and\ \citenamefont
  {Thomlinson}(2003)}]{Suortti2003MedicalRadiation}%
  \BibitemOpen
  \bibfield  {author} {\bibinfo {author} {\bibfnamefont {P.}~\bibnamefont
  {Suortti}}\ and\ \bibinfo {author} {\bibfnamefont {W.}~\bibnamefont
  {Thomlinson}},\ }\href {\doibase 10.1080/08940886.2011.567156} {\bibfield
  {journal} {\bibinfo  {journal} {Phys. Med. Biol.}\ }\textbf {\bibinfo
  {volume} {48}},\ \unskip\ \bibinfo {pages} {R1--R35} (\bibinfo {year}
  {2003})}\BibitemShut {NoStop}%
\bibitem [{\citenamefont {Chen}, \citenamefont {Bennett},\ and\ \citenamefont
  {Perticone}(2007)}]{Chen2007Dual-energyDetection}%
  \BibitemOpen
  \bibfield  {author} {\bibinfo {author} {\bibfnamefont {G.}~\bibnamefont
  {Chen}}, \bibinfo {author} {\bibfnamefont {G.}~\bibnamefont {Bennett}}, \
  and\ \bibinfo {author} {\bibfnamefont {D.}~\bibnamefont {Perticone}},\ }\href
  {\doibase 10.1016/j.nimb.2007.04.036} {\bibfield  {journal} {\bibinfo
  {journal} {Nuclear Instruments and Methods in Physics Research, Section B:
  Beam Interactions with Materials and Atoms}\ }\textbf {\bibinfo {volume}
  {261}},\ \unskip\ \bibinfo {pages} {356--359} (\bibinfo {year}
  {2007})}\BibitemShut {NoStop}%
\bibitem [{\citenamefont {Falk}\ \emph {et~al.}(2014)\citenamefont {Falk},
  \citenamefont {Collins}, \citenamefont {Gamboa}, \citenamefont {Kagan},
  \citenamefont {Kress}, \citenamefont {Montgomery}, \citenamefont
  {Srinivasan}, \citenamefont {Tzeferacos},\ and\ \citenamefont
  {Benage}}]{Falk2014CombinedShock-and-release}%
  \BibitemOpen
  \bibfield  {author} {\bibinfo {author} {\bibfnamefont {K.}~\bibnamefont
  {Falk}}, \bibinfo {author} {\bibfnamefont {L.~A.}\ \bibnamefont {Collins}},
  \bibinfo {author} {\bibfnamefont {E.~J.}\ \bibnamefont {Gamboa}}, \bibinfo
  {author} {\bibfnamefont {G.}~\bibnamefont {Kagan}}, \bibinfo {author}
  {\bibfnamefont {J.~D.}\ \bibnamefont {Kress}}, \bibinfo {author}
  {\bibfnamefont {D.~S.}\ \bibnamefont {Montgomery}}, \bibinfo {author}
  {\bibfnamefont {B.}~\bibnamefont {Srinivasan}}, \bibinfo {author}
  {\bibfnamefont {P.}~\bibnamefont {Tzeferacos}}, \ and\ \bibinfo {author}
  {\bibfnamefont {J.~F.}\ \bibnamefont {Benage}},\ }\href {\doibase
  10.1063/1.4876613} {\bibfield  {journal} {\bibinfo  {journal} {Physics of
  Plasmas}\ }\textbf {\bibinfo {volume} {21}},\ \ \bibinfo {pages} {056309}
  (\bibinfo {year} {2014})}\BibitemShut {NoStop}%
\bibitem [{\citenamefont {Tajima}\ and\ \citenamefont
  {Dawson}(1979)}]{Tajima1979LaserAccelerator}%
  \BibitemOpen
  \bibfield  {author} {\bibinfo {author} {\bibfnamefont {T.}~\bibnamefont
  {Tajima}}\ and\ \bibinfo {author} {\bibfnamefont {J.~M.}\ \bibnamefont
  {Dawson}},\ }\href {\doibase 10.1103/PhysRevLett.43.267} {\bibfield
  {journal} {\bibinfo  {journal} {Physical Review Letters}\ }\textbf {\bibinfo
  {volume} {43}},\ \unskip\ \bibinfo {pages} {267--270} (\bibinfo {year}
  {1979})}\BibitemShut {NoStop}%
\bibitem [{\citenamefont {Esarey}, \citenamefont {Schroeder},\ and\
  \citenamefont {Leemans}(2009)}]{Esarey2009PhysicsAccelerators}%
  \BibitemOpen
  \bibfield  {author} {\bibinfo {author} {\bibfnamefont {E.}~\bibnamefont
  {Esarey}}, \bibinfo {author} {\bibfnamefont {C.~B.}\ \bibnamefont
  {Schroeder}}, \ and\ \bibinfo {author} {\bibfnamefont {W.~P.}\ \bibnamefont
  {Leemans}},\ }\href {\doibase 10.1103/RevModPhys.81.1229} {\bibfield
  {journal} {\bibinfo  {journal} {Reviews of Modern Physics}\ }\textbf
  {\bibinfo {volume} {81}},\ \unskip\ \bibinfo {pages} {1229--1285} (\bibinfo
  {year} {2009})}\BibitemShut {NoStop}%
\bibitem [{\citenamefont {Wang}\ \emph {et~al.}(2013)\citenamefont {Wang},
  \citenamefont {Zgadzaj}, \citenamefont {Fazel}, \citenamefont {Li},
  \citenamefont {Yi}, \citenamefont {Zhang}, \citenamefont {Henderson},
  \citenamefont {Chang}, \citenamefont {Korzekwa}, \citenamefont {Tsai},
  \citenamefont {Pai}, \citenamefont {Quevedo}, \citenamefont {Dyer},
  \citenamefont {Gaul}, \citenamefont {Martinez}, \citenamefont {Bernstein},
  \citenamefont {Borger}, \citenamefont {Spinks}, \citenamefont {Donovan},
  \citenamefont {Khudik}, \citenamefont {Shvets}, \citenamefont {Ditmire},\
  and\ \citenamefont {Downer}}]{Wang2013Quasi-monoenergeticGeV}%
  \BibitemOpen
  \bibfield  {author} {\bibinfo {author} {\bibfnamefont {X.}~\bibnamefont
  {Wang}}, \bibinfo {author} {\bibfnamefont {R.}~\bibnamefont {Zgadzaj}},
  \bibinfo {author} {\bibfnamefont {N.}~\bibnamefont {Fazel}}, \bibinfo
  {author} {\bibfnamefont {Z.}~\bibnamefont {Li}}, \bibinfo {author}
  {\bibfnamefont {S.~A.}\ \bibnamefont {Yi}}, \bibinfo {author} {\bibfnamefont
  {X.}~\bibnamefont {Zhang}}, \bibinfo {author} {\bibfnamefont
  {W.}~\bibnamefont {Henderson}}, \bibinfo {author} {\bibfnamefont {Y.~Y.}\
  \bibnamefont {Chang}}, \bibinfo {author} {\bibfnamefont {R.}~\bibnamefont
  {Korzekwa}}, \bibinfo {author} {\bibfnamefont {H.~E.}\ \bibnamefont {Tsai}},
  \bibinfo {author} {\bibfnamefont {C.~H.}\ \bibnamefont {Pai}}, \bibinfo
  {author} {\bibfnamefont {H.}~\bibnamefont {Quevedo}}, \bibinfo {author}
  {\bibfnamefont {G.}~\bibnamefont {Dyer}}, \bibinfo {author} {\bibfnamefont
  {E.}~\bibnamefont {Gaul}}, \bibinfo {author} {\bibfnamefont {M.}~\bibnamefont
  {Martinez}}, \bibinfo {author} {\bibfnamefont {A.~C.}\ \bibnamefont
  {Bernstein}}, \bibinfo {author} {\bibfnamefont {T.}~\bibnamefont {Borger}},
  \bibinfo {author} {\bibfnamefont {M.}~\bibnamefont {Spinks}}, \bibinfo
  {author} {\bibfnamefont {M.}~\bibnamefont {Donovan}}, \bibinfo {author}
  {\bibfnamefont {V.}~\bibnamefont {Khudik}}, \bibinfo {author} {\bibfnamefont
  {G.}~\bibnamefont {Shvets}}, \bibinfo {author} {\bibfnamefont
  {T.}~\bibnamefont {Ditmire}}, \ and\ \bibinfo {author} {\bibfnamefont
  {M.~C.}\ \bibnamefont {Downer}},\ }\href {\doibase 10.1038/ncomms2988}
  {\bibfield  {journal} {\bibinfo  {journal} {Nature Communications}\ }\textbf
  {\bibinfo {volume} {4}},\ \ \bibinfo {pages} {1988} (\bibinfo {year}
  {2013})}\BibitemShut {NoStop}%
\bibitem [{\citenamefont {Gonsalves}\ \emph {et~al.}(2019)\citenamefont
  {Gonsalves}, \citenamefont {Nakamura}, \citenamefont {Daniels}, \citenamefont
  {Benedetti}, \citenamefont {Pieronek}, \citenamefont {De~Raadt},
  \citenamefont {Steinke}, \citenamefont {Bin}, \citenamefont {Bulanov},
  \citenamefont {Van~Tilborg}, \citenamefont {Geddes}, \citenamefont
  {Schroeder}, \citenamefont {T{\'{o}}th}, \citenamefont {Esarey},
  \citenamefont {Swanson}, \citenamefont {Fan-Chiang}, \citenamefont
  {Bagdasarov}, \citenamefont {Bobrova}, \citenamefont {Gasilov}, \citenamefont
  {Korn}, \citenamefont {Sasorov},\ and\ \citenamefont
  {Leemans}}]{Gonsalves2019PetawattWaveguide}%
  \BibitemOpen
  \bibfield  {author} {\bibinfo {author} {\bibfnamefont {A.~J.}\ \bibnamefont
  {Gonsalves}}, \bibinfo {author} {\bibfnamefont {K.}~\bibnamefont {Nakamura}},
  \bibinfo {author} {\bibfnamefont {J.}~\bibnamefont {Daniels}}, \bibinfo
  {author} {\bibfnamefont {C.}~\bibnamefont {Benedetti}}, \bibinfo {author}
  {\bibfnamefont {C.}~\bibnamefont {Pieronek}}, \bibinfo {author}
  {\bibfnamefont {T.~C.}\ \bibnamefont {De~Raadt}}, \bibinfo {author}
  {\bibfnamefont {S.}~\bibnamefont {Steinke}}, \bibinfo {author} {\bibfnamefont
  {J.~H.}\ \bibnamefont {Bin}}, \bibinfo {author} {\bibfnamefont {S.~S.}\
  \bibnamefont {Bulanov}}, \bibinfo {author} {\bibfnamefont {J.}~\bibnamefont
  {Van~Tilborg}}, \bibinfo {author} {\bibfnamefont {C.~G.}\ \bibnamefont
  {Geddes}}, \bibinfo {author} {\bibfnamefont {C.~B.}\ \bibnamefont
  {Schroeder}}, \bibinfo {author} {\bibfnamefont {C.}~\bibnamefont
  {T{\'{o}}th}}, \bibinfo {author} {\bibfnamefont {E.}~\bibnamefont {Esarey}},
  \bibinfo {author} {\bibfnamefont {K.}~\bibnamefont {Swanson}}, \bibinfo
  {author} {\bibfnamefont {L.}~\bibnamefont {Fan-Chiang}}, \bibinfo {author}
  {\bibfnamefont {G.}~\bibnamefont {Bagdasarov}}, \bibinfo {author}
  {\bibfnamefont {N.}~\bibnamefont {Bobrova}}, \bibinfo {author} {\bibfnamefont
  {V.}~\bibnamefont {Gasilov}}, \bibinfo {author} {\bibfnamefont
  {G.}~\bibnamefont {Korn}}, \bibinfo {author} {\bibfnamefont {P.}~\bibnamefont
  {Sasorov}}, \ and\ \bibinfo {author} {\bibfnamefont {W.~P.}\ \bibnamefont
  {Leemans}},\ }\href {\doibase 10.1103/PhysRevLett.122.084801} {\bibfield
  {journal} {\bibinfo  {journal} {Physical Review Letters}\ }\textbf {\bibinfo
  {volume} {122}},\ \ \bibinfo {pages} {084801} (\bibinfo {year}
  {2019})}\BibitemShut {NoStop}%
\bibitem [{\citenamefont {Couperus}\ \emph {et~al.}(2017)\citenamefont
  {Couperus}, \citenamefont {Pausch}, \citenamefont {K{\"{o}}hler},
  \citenamefont {Zarini}, \citenamefont {Kr{\"{a}}mer}, \citenamefont {Garten},
  \citenamefont {Huebl}, \citenamefont {Gebhardt}, \citenamefont {Helbig},
  \citenamefont {Bock}, \citenamefont {Zeil}, \citenamefont {Debus},
  \citenamefont {Bussmann}, \citenamefont {Schramm},\ and\ \citenamefont
  {Irman}}]{Couperus2017DemonstrationAccelerator}%
  \BibitemOpen
  \bibfield  {author} {\bibinfo {author} {\bibfnamefont {J.~P.}\ \bibnamefont
  {Couperus}}, \bibinfo {author} {\bibfnamefont {R.}~\bibnamefont {Pausch}},
  \bibinfo {author} {\bibfnamefont {A.}~\bibnamefont {K{\"{o}}hler}}, \bibinfo
  {author} {\bibfnamefont {O.}~\bibnamefont {Zarini}}, \bibinfo {author}
  {\bibfnamefont {J.~M.}\ \bibnamefont {Kr{\"{a}}mer}}, \bibinfo {author}
  {\bibfnamefont {M.}~\bibnamefont {Garten}}, \bibinfo {author} {\bibfnamefont
  {A.}~\bibnamefont {Huebl}}, \bibinfo {author} {\bibfnamefont
  {R.}~\bibnamefont {Gebhardt}}, \bibinfo {author} {\bibfnamefont
  {U.}~\bibnamefont {Helbig}}, \bibinfo {author} {\bibfnamefont
  {S.}~\bibnamefont {Bock}}, \bibinfo {author} {\bibfnamefont {K.}~\bibnamefont
  {Zeil}}, \bibinfo {author} {\bibfnamefont {A.}~\bibnamefont {Debus}},
  \bibinfo {author} {\bibfnamefont {M.}~\bibnamefont {Bussmann}}, \bibinfo
  {author} {\bibfnamefont {U.}~\bibnamefont {Schramm}}, \ and\ \bibinfo
  {author} {\bibfnamefont {A.}~\bibnamefont {Irman}},\ }\href {\doibase
  10.1038/s41467-017-00592-7} {\bibfield  {journal} {\bibinfo  {journal}
  {Nature Communications}\ }\textbf {\bibinfo {volume} {8}},\ \ \bibinfo
  {pages} {487} (\bibinfo {year} {2017})}\BibitemShut {NoStop}%
\bibitem [{\citenamefont {G{\"{o}}tzfried}\ \emph {et~al.}(2020)\citenamefont
  {G{\"{o}}tzfried}, \citenamefont {D{\"{o}}pp}, \citenamefont {Gilljohann},
  \citenamefont {Foerster}, \citenamefont {Ding}, \citenamefont {Schindler},
  \citenamefont {Schilling}, \citenamefont {Buck}, \citenamefont {Veisz},\ and\
  \citenamefont {Karsch}}]{Gotzfried2020PhysicsWakefields}%
  \BibitemOpen
  \bibfield  {author} {\bibinfo {author} {\bibfnamefont {J.}~\bibnamefont
  {G{\"{o}}tzfried}}, \bibinfo {author} {\bibfnamefont {A.}~\bibnamefont
  {D{\"{o}}pp}}, \bibinfo {author} {\bibfnamefont {M.~F.}\ \bibnamefont
  {Gilljohann}}, \bibinfo {author} {\bibfnamefont {F.~M.}\ \bibnamefont
  {Foerster}}, \bibinfo {author} {\bibfnamefont {H.}~\bibnamefont {Ding}},
  \bibinfo {author} {\bibfnamefont {S.}~\bibnamefont {Schindler}}, \bibinfo
  {author} {\bibfnamefont {G.}~\bibnamefont {Schilling}}, \bibinfo {author}
  {\bibfnamefont {A.}~\bibnamefont {Buck}}, \bibinfo {author} {\bibfnamefont
  {L.}~\bibnamefont {Veisz}}, \ and\ \bibinfo {author} {\bibfnamefont
  {S.}~\bibnamefont {Karsch}},\ }\href {\doibase 10.1103/PhysRevX.10.041015}
  {\bibfield  {journal} {\bibinfo  {journal} {Physical Review X}\ }\textbf
  {\bibinfo {volume} {10}},\ \ \bibinfo {pages} {041015} (\bibinfo {year}
  {2020})}\BibitemShut {NoStop}%
\bibitem [{\citenamefont {Corde}\ \emph {et~al.}(2013)\citenamefont {Corde},
  \citenamefont {Ta~Phuoc}, \citenamefont {Lambert}, \citenamefont {Fitour},
  \citenamefont {Malka}, \citenamefont {Rousse}, \citenamefont {Beck},\ and\
  \citenamefont {Lefebvre}}]{Corde2013FemtosecondAccelerators}%
  \BibitemOpen
  \bibfield  {author} {\bibinfo {author} {\bibfnamefont {S.}~\bibnamefont
  {Corde}}, \bibinfo {author} {\bibfnamefont {K.}~\bibnamefont {Ta~Phuoc}},
  \bibinfo {author} {\bibfnamefont {G.}~\bibnamefont {Lambert}}, \bibinfo
  {author} {\bibfnamefont {R.}~\bibnamefont {Fitour}}, \bibinfo {author}
  {\bibfnamefont {V.}~\bibnamefont {Malka}}, \bibinfo {author} {\bibfnamefont
  {A.}~\bibnamefont {Rousse}}, \bibinfo {author} {\bibfnamefont
  {A.}~\bibnamefont {Beck}}, \ and\ \bibinfo {author} {\bibfnamefont
  {E.}~\bibnamefont {Lefebvre}},\ }\href {\doibase 10.1103/RevModPhys.85.1}
  {\bibfield  {journal} {\bibinfo  {journal} {Reviews of Modern Physics}\
  }\textbf {\bibinfo {volume} {85}},\ \unskip\ \bibinfo {pages} {1--48}
  (\bibinfo {year} {2013})}\BibitemShut {NoStop}%
\bibitem [{\citenamefont {Kr{\"{a}}mer}\ \emph {et~al.}(2018)\citenamefont
  {Kr{\"{a}}mer}, \citenamefont {Jochmann}, \citenamefont {Budde},
  \citenamefont {Bussmann}, \citenamefont {Couperus}, \citenamefont {Cowan},
  \citenamefont {Debus}, \citenamefont {K{\"{o}}hler}, \citenamefont
  {Kuntzsch}, \citenamefont {Laso~Garc{\'{i}}a}, \citenamefont {Lehnert},
  \citenamefont {Michel}, \citenamefont {Pausch}, \citenamefont {Zarini},
  \citenamefont {Schramm},\ and\ \citenamefont
  {Irman}}]{Kramer2018MakingApplications}%
  \BibitemOpen
  \bibfield  {author} {\bibinfo {author} {\bibfnamefont {J.~M.}\ \bibnamefont
  {Kr{\"{a}}mer}}, \bibinfo {author} {\bibfnamefont {A.}~\bibnamefont
  {Jochmann}}, \bibinfo {author} {\bibfnamefont {M.}~\bibnamefont {Budde}},
  \bibinfo {author} {\bibfnamefont {M.}~\bibnamefont {Bussmann}}, \bibinfo
  {author} {\bibfnamefont {J.~P.}\ \bibnamefont {Couperus}}, \bibinfo {author}
  {\bibfnamefont {T.~E.}\ \bibnamefont {Cowan}}, \bibinfo {author}
  {\bibfnamefont {A.}~\bibnamefont {Debus}}, \bibinfo {author} {\bibfnamefont
  {A.}~\bibnamefont {K{\"{o}}hler}}, \bibinfo {author} {\bibfnamefont
  {M.}~\bibnamefont {Kuntzsch}}, \bibinfo {author} {\bibfnamefont
  {A.}~\bibnamefont {Laso~Garc{\'{i}}a}}, \bibinfo {author} {\bibfnamefont
  {U.}~\bibnamefont {Lehnert}}, \bibinfo {author} {\bibfnamefont
  {P.}~\bibnamefont {Michel}}, \bibinfo {author} {\bibfnamefont
  {R.}~\bibnamefont {Pausch}}, \bibinfo {author} {\bibfnamefont
  {O.}~\bibnamefont {Zarini}}, \bibinfo {author} {\bibfnamefont
  {U.}~\bibnamefont {Schramm}}, \ and\ \bibinfo {author} {\bibfnamefont
  {A.}~\bibnamefont {Irman}},\ }\href {\doibase 10.1038/s41598-018-19546-0}
  {\bibfield  {journal} {\bibinfo  {journal} {Scientific Reports}\ }\textbf
  {\bibinfo {volume} {8}},\ \unskip\ \bibinfo {pages} {1--11} (\bibinfo {year}
  {2018})}\BibitemShut {NoStop}%
\bibitem [{\citenamefont {Khrennikov}\ \emph {et~al.}(2015)\citenamefont
  {Khrennikov}, \citenamefont {Wenz}, \citenamefont {Buck}, \citenamefont {Xu},
  \citenamefont {Heigoldt}, \citenamefont {Veisz},\ and\ \citenamefont
  {Karsch}}]{Khrennikov2015TunableRegime}%
  \BibitemOpen
  \bibfield  {author} {\bibinfo {author} {\bibfnamefont {K.}~\bibnamefont
  {Khrennikov}}, \bibinfo {author} {\bibfnamefont {J.}~\bibnamefont {Wenz}},
  \bibinfo {author} {\bibfnamefont {A.}~\bibnamefont {Buck}}, \bibinfo {author}
  {\bibfnamefont {J.}~\bibnamefont {Xu}}, \bibinfo {author} {\bibfnamefont
  {M.}~\bibnamefont {Heigoldt}}, \bibinfo {author} {\bibfnamefont
  {L.}~\bibnamefont {Veisz}}, \ and\ \bibinfo {author} {\bibfnamefont
  {S.}~\bibnamefont {Karsch}},\ }\href {\doibase
  10.1103/PhysRevLett.114.195003} {\bibfield  {journal} {\bibinfo  {journal}
  {Physical Review Letters}\ }\textbf {\bibinfo {volume} {114}},\ \ \bibinfo
  {pages} {195003} (\bibinfo {year} {2015})}\BibitemShut {NoStop}%
\bibitem [{\citenamefont {Yan}\ \emph {et~al.}(2017)\citenamefont {Yan},
  \citenamefont {Fruhling}, \citenamefont {Golovin}, \citenamefont {Haden},
  \citenamefont {Luo}, \citenamefont {Zhang}, \citenamefont {Zhao},
  \citenamefont {Zhang}, \citenamefont {Liu}, \citenamefont {Chen},
  \citenamefont {Chen}, \citenamefont {Banerjee},\ and\ \citenamefont
  {Umstadter}}]{Yan2017High-orderScattering}%
  \BibitemOpen
  \bibfield  {author} {\bibinfo {author} {\bibfnamefont {W.}~\bibnamefont
  {Yan}}, \bibinfo {author} {\bibfnamefont {C.}~\bibnamefont {Fruhling}},
  \bibinfo {author} {\bibfnamefont {G.}~\bibnamefont {Golovin}}, \bibinfo
  {author} {\bibfnamefont {D.}~\bibnamefont {Haden}}, \bibinfo {author}
  {\bibfnamefont {J.}~\bibnamefont {Luo}}, \bibinfo {author} {\bibfnamefont
  {P.}~\bibnamefont {Zhang}}, \bibinfo {author} {\bibfnamefont
  {B.}~\bibnamefont {Zhao}}, \bibinfo {author} {\bibfnamefont {J.}~\bibnamefont
  {Zhang}}, \bibinfo {author} {\bibfnamefont {C.}~\bibnamefont {Liu}}, \bibinfo
  {author} {\bibfnamefont {M.}~\bibnamefont {Chen}}, \bibinfo {author}
  {\bibfnamefont {S.}~\bibnamefont {Chen}}, \bibinfo {author} {\bibfnamefont
  {S.}~\bibnamefont {Banerjee}}, \ and\ \bibinfo {author} {\bibfnamefont
  {D.}~\bibnamefont {Umstadter}},\ }\href {\doibase 10.1038/nphoton.2017.100}
  {\bibfield  {journal} {\bibinfo  {journal} {Nature Photonics}\ }\textbf
  {\bibinfo {volume} {11}},\ \unskip\ \bibinfo {pages} {514--521} (\bibinfo
  {year} {2017})}\BibitemShut {NoStop}%
\bibitem [{\citenamefont {Cole}\ \emph {et~al.}(2018)\citenamefont {Cole},
  \citenamefont {Behm}, \citenamefont {Gerstmayr}, \citenamefont {Blackburn},
  \citenamefont {Wood}, \citenamefont {Baird}, \citenamefont {Duff},
  \citenamefont {Harvey}, \citenamefont {Ilderton}, \citenamefont {Joglekar},
  \citenamefont {Krushelnick}, \citenamefont {Kuschel}, \citenamefont
  {Marklund}, \citenamefont {McKenna}, \citenamefont {Murphy}, \citenamefont
  {Poder}, \citenamefont {Ridgers}, \citenamefont {Samarin}, \citenamefont
  {Sarri}, \citenamefont {Symes}, \citenamefont {Thomas}, \citenamefont
  {Warwick}, \citenamefont {Zepf}, \citenamefont {Najmudin},\ and\
  \citenamefont {Mangles}}]{Cole2018ExperimentalBeam}%
  \BibitemOpen
  \bibfield  {author} {\bibinfo {author} {\bibfnamefont {J.~M.}\ \bibnamefont
  {Cole}}, \bibinfo {author} {\bibfnamefont {K.~T.}\ \bibnamefont {Behm}},
  \bibinfo {author} {\bibfnamefont {E.}~\bibnamefont {Gerstmayr}}, \bibinfo
  {author} {\bibfnamefont {T.~G.}\ \bibnamefont {Blackburn}}, \bibinfo {author}
  {\bibfnamefont {J.~C.}\ \bibnamefont {Wood}}, \bibinfo {author}
  {\bibfnamefont {C.~D.}\ \bibnamefont {Baird}}, \bibinfo {author}
  {\bibfnamefont {M.~J.}\ \bibnamefont {Duff}}, \bibinfo {author}
  {\bibfnamefont {C.}~\bibnamefont {Harvey}}, \bibinfo {author} {\bibfnamefont
  {A.}~\bibnamefont {Ilderton}}, \bibinfo {author} {\bibfnamefont {A.~S.}\
  \bibnamefont {Joglekar}}, \bibinfo {author} {\bibfnamefont {K.}~\bibnamefont
  {Krushelnick}}, \bibinfo {author} {\bibfnamefont {S.}~\bibnamefont
  {Kuschel}}, \bibinfo {author} {\bibfnamefont {M.}~\bibnamefont {Marklund}},
  \bibinfo {author} {\bibfnamefont {P.}~\bibnamefont {McKenna}}, \bibinfo
  {author} {\bibfnamefont {C.~D.}\ \bibnamefont {Murphy}}, \bibinfo {author}
  {\bibfnamefont {K.}~\bibnamefont {Poder}}, \bibinfo {author} {\bibfnamefont
  {C.~P.}\ \bibnamefont {Ridgers}}, \bibinfo {author} {\bibfnamefont {G.~M.}\
  \bibnamefont {Samarin}}, \bibinfo {author} {\bibfnamefont {G.}~\bibnamefont
  {Sarri}}, \bibinfo {author} {\bibfnamefont {D.~R.}\ \bibnamefont {Symes}},
  \bibinfo {author} {\bibfnamefont {A.~G.}\ \bibnamefont {Thomas}}, \bibinfo
  {author} {\bibfnamefont {J.}~\bibnamefont {Warwick}}, \bibinfo {author}
  {\bibfnamefont {M.}~\bibnamefont {Zepf}}, \bibinfo {author} {\bibfnamefont
  {Z.}~\bibnamefont {Najmudin}}, \ and\ \bibinfo {author} {\bibfnamefont
  {S.~P.}\ \bibnamefont {Mangles}},\ }\href {\doibase
  10.1103/PhysRevX.8.011020} {\bibfield  {journal} {\bibinfo  {journal}
  {Physical Review X}\ }\textbf {\bibinfo {volume} {8}},\ \ \bibinfo {pages}
  {011020} (\bibinfo {year} {2018})}\BibitemShut {NoStop}%
\bibitem [{\citenamefont {Sarri}\ \emph {et~al.}(2014)\citenamefont {Sarri},
  \citenamefont {Corvan}, \citenamefont {Schumaker}, \citenamefont {Cole},
  \citenamefont {Di~Piazza}, \citenamefont {Ahmed}, \citenamefont {Harvey},
  \citenamefont {Keitel}, \citenamefont {Krushelnick}, \citenamefont {Mangles},
  \citenamefont {Najmudin}, \citenamefont {Symes}, \citenamefont {Thomas},
  \citenamefont {Yeung}, \citenamefont {Zhao},\ and\ \citenamefont
  {Zepf}}]{Sarri2014UltrahighScattering}%
  \BibitemOpen
  \bibfield  {author} {\bibinfo {author} {\bibfnamefont {G.}~\bibnamefont
  {Sarri}}, \bibinfo {author} {\bibfnamefont {D.~J.}\ \bibnamefont {Corvan}},
  \bibinfo {author} {\bibfnamefont {W.}~\bibnamefont {Schumaker}}, \bibinfo
  {author} {\bibfnamefont {J.~M.}\ \bibnamefont {Cole}}, \bibinfo {author}
  {\bibfnamefont {A.}~\bibnamefont {Di~Piazza}}, \bibinfo {author}
  {\bibfnamefont {H.}~\bibnamefont {Ahmed}}, \bibinfo {author} {\bibfnamefont
  {C.}~\bibnamefont {Harvey}}, \bibinfo {author} {\bibfnamefont {C.~H.}\
  \bibnamefont {Keitel}}, \bibinfo {author} {\bibfnamefont {K.}~\bibnamefont
  {Krushelnick}}, \bibinfo {author} {\bibfnamefont {S.~P.}\ \bibnamefont
  {Mangles}}, \bibinfo {author} {\bibfnamefont {Z.}~\bibnamefont {Najmudin}},
  \bibinfo {author} {\bibfnamefont {D.}~\bibnamefont {Symes}}, \bibinfo
  {author} {\bibfnamefont {A.~G.}\ \bibnamefont {Thomas}}, \bibinfo {author}
  {\bibfnamefont {M.}~\bibnamefont {Yeung}}, \bibinfo {author} {\bibfnamefont
  {Z.}~\bibnamefont {Zhao}}, \ and\ \bibinfo {author} {\bibfnamefont
  {M.}~\bibnamefont {Zepf}},\ }\href {\doibase 10.1103/PhysRevLett.113.224801}
  {\bibfield  {journal} {\bibinfo  {journal} {Physical Review Letters}\
  }\textbf {\bibinfo {volume} {113}},\ \ \bibinfo {pages} {224801} (\bibinfo
  {year} {2014})}\BibitemShut {NoStop}%
\bibitem [{\citenamefont {Ta~Phuoc}\ \emph {et~al.}(2012)\citenamefont
  {Ta~Phuoc}, \citenamefont {Corde}, \citenamefont {Thaury}, \citenamefont
  {Malka}, \citenamefont {Tafzi}, \citenamefont {Goddet}, \citenamefont {Shah},
  \citenamefont {Sebban},\ and\ \citenamefont
  {Rousse}}]{TaPhuoc2012All-opticalSource}%
  \BibitemOpen
  \bibfield  {author} {\bibinfo {author} {\bibfnamefont {K.}~\bibnamefont
  {Ta~Phuoc}}, \bibinfo {author} {\bibfnamefont {S.}~\bibnamefont {Corde}},
  \bibinfo {author} {\bibfnamefont {C.}~\bibnamefont {Thaury}}, \bibinfo
  {author} {\bibfnamefont {V.}~\bibnamefont {Malka}}, \bibinfo {author}
  {\bibfnamefont {A.}~\bibnamefont {Tafzi}}, \bibinfo {author} {\bibfnamefont
  {J.~P.}\ \bibnamefont {Goddet}}, \bibinfo {author} {\bibfnamefont {R.~C.}\
  \bibnamefont {Shah}}, \bibinfo {author} {\bibfnamefont {S.}~\bibnamefont
  {Sebban}}, \ and\ \bibinfo {author} {\bibfnamefont {A.}~\bibnamefont
  {Rousse}},\ }\href {\doibase 10.1038/nphoton.2012.82} {\bibfield  {journal}
  {\bibinfo  {journal} {Nature Photonics}\ }\textbf {\bibinfo {volume} {6}},\
  \unskip\ \bibinfo {pages} {308--311} (\bibinfo {year} {2012})}\BibitemShut
  {NoStop}%
\bibitem [{\citenamefont {Tsai}\ \emph {et~al.}(2015)\citenamefont {Tsai},
  \citenamefont {Wang}, \citenamefont {Shaw}, \citenamefont {Arefiev},
  \citenamefont {Li}, \citenamefont {Zhang}, \citenamefont {Zgadzaj},
  \citenamefont {Henderson}, \citenamefont {Khudik}, \citenamefont {Shvets},\
  and\ \citenamefont {Downer}}]{Tsai2015CompactMirror}%
  \BibitemOpen
  \bibfield  {author} {\bibinfo {author} {\bibfnamefont {H.~E.}\ \bibnamefont
  {Tsai}}, \bibinfo {author} {\bibfnamefont {X.}~\bibnamefont {Wang}}, \bibinfo
  {author} {\bibfnamefont {J.}~\bibnamefont {Shaw}}, \bibinfo {author}
  {\bibfnamefont {A.~V.}\ \bibnamefont {Arefiev}}, \bibinfo {author}
  {\bibfnamefont {Z.}~\bibnamefont {Li}}, \bibinfo {author} {\bibfnamefont
  {X.}~\bibnamefont {Zhang}}, \bibinfo {author} {\bibfnamefont
  {R.}~\bibnamefont {Zgadzaj}}, \bibinfo {author} {\bibfnamefont
  {W.}~\bibnamefont {Henderson}}, \bibinfo {author} {\bibfnamefont
  {V.}~\bibnamefont {Khudik}}, \bibinfo {author} {\bibfnamefont
  {G.}~\bibnamefont {Shvets}}, \ and\ \bibinfo {author} {\bibfnamefont {M.~C.}\
  \bibnamefont {Downer}},\ }\href {\doibase 10.1063/1.4965663} {\bibfield
  {journal} {\bibinfo  {journal} {Physics of Plasmas}\ }\textbf {\bibinfo
  {volume} {22}},\ \ \bibinfo {pages} {023106} (\bibinfo {year}
  {2015})}\BibitemShut {NoStop}%
\bibitem [{\citenamefont {Irman}\ \emph {et~al.}(2018)\citenamefont {Irman},
  \citenamefont {Couperus}, \citenamefont {Debus}, \citenamefont
  {K{\"{o}}hler}, \citenamefont {Kr{\"{a}}mer}, \citenamefont {Pausch},
  \citenamefont {Zarini},\ and\ \citenamefont
  {Schramm}}]{Irman2018ImprovedInjection}%
  \BibitemOpen
  \bibfield  {author} {\bibinfo {author} {\bibfnamefont {A.}~\bibnamefont
  {Irman}}, \bibinfo {author} {\bibfnamefont {J.~P.}\ \bibnamefont {Couperus}},
  \bibinfo {author} {\bibfnamefont {A.}~\bibnamefont {Debus}}, \bibinfo
  {author} {\bibfnamefont {A.}~\bibnamefont {K{\"{o}}hler}}, \bibinfo {author}
  {\bibfnamefont {J.~M.}\ \bibnamefont {Kr{\"{a}}mer}}, \bibinfo {author}
  {\bibfnamefont {R.}~\bibnamefont {Pausch}}, \bibinfo {author} {\bibfnamefont
  {O.}~\bibnamefont {Zarini}}, \ and\ \bibinfo {author} {\bibfnamefont
  {U.}~\bibnamefont {Schramm}},\ }\href {\doibase 10.1088/1361-6587/aaaef1}
  {\bibfield  {journal} {\bibinfo  {journal} {Plasma Physics and Controlled
  Fusion}\ }\textbf {\bibinfo {volume} {60}},\ \ \bibinfo {pages} {044015}
  (\bibinfo {year} {2018})}\BibitemShut {NoStop}%
\bibitem [{\citenamefont {Mirzaie}\ \emph {et~al.}(2015)\citenamefont
  {Mirzaie}, \citenamefont {Li}, \citenamefont {Zeng}, \citenamefont {Hafz},
  \citenamefont {Chen}, \citenamefont {Li}, \citenamefont {Zhu}, \citenamefont
  {Liao}, \citenamefont {Sokollik}, \citenamefont {Liu}, \citenamefont {Ma},
  \citenamefont {Chen}, \citenamefont {Sheng},\ and\ \citenamefont
  {Zhang}}]{Mirzaie2015DemonstrationBeams}%
  \BibitemOpen
  \bibfield  {author} {\bibinfo {author} {\bibfnamefont {M.}~\bibnamefont
  {Mirzaie}}, \bibinfo {author} {\bibfnamefont {S.}~\bibnamefont {Li}},
  \bibinfo {author} {\bibfnamefont {M.}~\bibnamefont {Zeng}}, \bibinfo {author}
  {\bibfnamefont {N.~A.}\ \bibnamefont {Hafz}}, \bibinfo {author}
  {\bibfnamefont {M.}~\bibnamefont {Chen}}, \bibinfo {author} {\bibfnamefont
  {G.~Y.}\ \bibnamefont {Li}}, \bibinfo {author} {\bibfnamefont {Q.~J.}\
  \bibnamefont {Zhu}}, \bibinfo {author} {\bibfnamefont {H.}~\bibnamefont
  {Liao}}, \bibinfo {author} {\bibfnamefont {T.}~\bibnamefont {Sokollik}},
  \bibinfo {author} {\bibfnamefont {F.}~\bibnamefont {Liu}}, \bibinfo {author}
  {\bibfnamefont {Y.~Y.}\ \bibnamefont {Ma}}, \bibinfo {author} {\bibfnamefont
  {L.~M.}\ \bibnamefont {Chen}}, \bibinfo {author} {\bibfnamefont {Z.~M.}\
  \bibnamefont {Sheng}}, \ and\ \bibinfo {author} {\bibfnamefont
  {J.}~\bibnamefont {Zhang}},\ }\href {\doibase 10.1038/srep14659} {\bibfield
  {journal} {\bibinfo  {journal} {Scientific Reports}\ }\textbf {\bibinfo
  {volume} {5}},\ \unskip\ \bibinfo {pages} {1--9} (\bibinfo {year}
  {2015})}\BibitemShut {NoStop}%
\bibitem [{\citenamefont {Schramm}\ \emph {et~al.}(2017)\citenamefont
  {Schramm}, \citenamefont {Bussmann}, \citenamefont {Irman}, \citenamefont
  {Siebold}, \citenamefont {Zeil}, \citenamefont {Albach}, \citenamefont
  {Bernert}, \citenamefont {Bock}, \citenamefont {Brack}, \citenamefont
  {Branco}, \citenamefont {Couperus}, \citenamefont {Cowan}, \citenamefont
  {Debus}, \citenamefont {Eisenmann}, \citenamefont {Garten}, \citenamefont
  {Gebhardt}, \citenamefont {Grams}, \citenamefont {Helbig}, \citenamefont
  {Huebl}, \citenamefont {Kluge}, \citenamefont {K{\"{o}}hler}, \citenamefont
  {Kr{\"{a}}mer}, \citenamefont {Kraft}, \citenamefont {Kroll}, \citenamefont
  {Kuntzsch}, \citenamefont {Lehnert}, \citenamefont {Loeser}, \citenamefont
  {Metzkes}, \citenamefont {Michel}, \citenamefont {Obst}, \citenamefont
  {Pausch}, \citenamefont {Rehwald}, \citenamefont {Sauerbrey}, \citenamefont
  {Schlenvoigt}, \citenamefont {Steiniger},\ and\ \citenamefont
  {Zarini}}]{Schramm2017FirstDresden}%
  \BibitemOpen
  \bibfield  {author} {\bibinfo {author} {\bibfnamefont {U.}~\bibnamefont
  {Schramm}}, \bibinfo {author} {\bibfnamefont {M.}~\bibnamefont {Bussmann}},
  \bibinfo {author} {\bibfnamefont {A.}~\bibnamefont {Irman}}, \bibinfo
  {author} {\bibfnamefont {M.}~\bibnamefont {Siebold}}, \bibinfo {author}
  {\bibfnamefont {K.}~\bibnamefont {Zeil}}, \bibinfo {author} {\bibfnamefont
  {D.}~\bibnamefont {Albach}}, \bibinfo {author} {\bibfnamefont
  {C.}~\bibnamefont {Bernert}}, \bibinfo {author} {\bibfnamefont
  {S.}~\bibnamefont {Bock}}, \bibinfo {author} {\bibfnamefont {F.}~\bibnamefont
  {Brack}}, \bibinfo {author} {\bibfnamefont {J.}~\bibnamefont {Branco}},
  \bibinfo {author} {\bibfnamefont {J.~P.}\ \bibnamefont {Couperus}}, \bibinfo
  {author} {\bibfnamefont {T.~E.}\ \bibnamefont {Cowan}}, \bibinfo {author}
  {\bibfnamefont {A.}~\bibnamefont {Debus}}, \bibinfo {author} {\bibfnamefont
  {C.}~\bibnamefont {Eisenmann}}, \bibinfo {author} {\bibfnamefont
  {M.}~\bibnamefont {Garten}}, \bibinfo {author} {\bibfnamefont
  {R.}~\bibnamefont {Gebhardt}}, \bibinfo {author} {\bibfnamefont
  {S.}~\bibnamefont {Grams}}, \bibinfo {author} {\bibfnamefont
  {U.}~\bibnamefont {Helbig}}, \bibinfo {author} {\bibfnamefont
  {A.}~\bibnamefont {Huebl}}, \bibinfo {author} {\bibfnamefont
  {T.}~\bibnamefont {Kluge}}, \bibinfo {author} {\bibfnamefont
  {A.}~\bibnamefont {K{\"{o}}hler}}, \bibinfo {author} {\bibfnamefont {J.~M.}\
  \bibnamefont {Kr{\"{a}}mer}}, \bibinfo {author} {\bibfnamefont
  {S.}~\bibnamefont {Kraft}}, \bibinfo {author} {\bibfnamefont
  {F.}~\bibnamefont {Kroll}}, \bibinfo {author} {\bibfnamefont
  {M.}~\bibnamefont {Kuntzsch}}, \bibinfo {author} {\bibfnamefont
  {U.}~\bibnamefont {Lehnert}}, \bibinfo {author} {\bibfnamefont
  {M.}~\bibnamefont {Loeser}}, \bibinfo {author} {\bibfnamefont
  {J.}~\bibnamefont {Metzkes}}, \bibinfo {author} {\bibfnamefont
  {P.}~\bibnamefont {Michel}}, \bibinfo {author} {\bibfnamefont
  {L.}~\bibnamefont {Obst}}, \bibinfo {author} {\bibfnamefont {R.}~\bibnamefont
  {Pausch}}, \bibinfo {author} {\bibfnamefont {M.}~\bibnamefont {Rehwald}},
  \bibinfo {author} {\bibfnamefont {R.}~\bibnamefont {Sauerbrey}}, \bibinfo
  {author} {\bibfnamefont {H.~P.}\ \bibnamefont {Schlenvoigt}}, \bibinfo
  {author} {\bibfnamefont {K.}~\bibnamefont {Steiniger}}, \ and\ \bibinfo
  {author} {\bibfnamefont {O.}~\bibnamefont {Zarini}},\ }\href {\doibase
  10.1088/1742-6596/874/1/012028} {\bibfield  {journal} {\bibinfo  {journal}
  {Journal of Physics: Conference Series}\ }\textbf {\bibinfo {volume} {874}},\
  \ \bibinfo {pages} {012028} (\bibinfo {year} {2017})}\BibitemShut {NoStop}%
\bibitem [{\citenamefont {Kurz}\ \emph {et~al.}(2018)\citenamefont {Kurz},
  \citenamefont {Couperus}, \citenamefont {Kr{\"{a}}mer}, \citenamefont {Ding},
  \citenamefont {Kuschel}, \citenamefont {K{\"{o}}hler}, \citenamefont
  {Zarini}, \citenamefont {Hollatz}, \citenamefont {Schinkel}, \citenamefont
  {D'Arcy}, \citenamefont {Schwinkendorf}, \citenamefont {Osterhoff},
  \citenamefont {Irman}, \citenamefont {Schramm},\ and\ \citenamefont
  {Karsch}}]{Kurz2018CalibrationDetermination}%
  \BibitemOpen
  \bibfield  {author} {\bibinfo {author} {\bibfnamefont {T.}~\bibnamefont
  {Kurz}}, \bibinfo {author} {\bibfnamefont {J.~P.}\ \bibnamefont {Couperus}},
  \bibinfo {author} {\bibfnamefont {J.~M.}\ \bibnamefont {Kr{\"{a}}mer}},
  \bibinfo {author} {\bibfnamefont {H.}~\bibnamefont {Ding}}, \bibinfo {author}
  {\bibfnamefont {S.}~\bibnamefont {Kuschel}}, \bibinfo {author} {\bibfnamefont
  {A.}~\bibnamefont {K{\"{o}}hler}}, \bibinfo {author} {\bibfnamefont
  {O.}~\bibnamefont {Zarini}}, \bibinfo {author} {\bibfnamefont
  {D.}~\bibnamefont {Hollatz}}, \bibinfo {author} {\bibfnamefont
  {D.}~\bibnamefont {Schinkel}}, \bibinfo {author} {\bibfnamefont
  {R.}~\bibnamefont {D'Arcy}}, \bibinfo {author} {\bibfnamefont {J.~P.}\
  \bibnamefont {Schwinkendorf}}, \bibinfo {author} {\bibfnamefont
  {J.}~\bibnamefont {Osterhoff}}, \bibinfo {author} {\bibfnamefont
  {A.}~\bibnamefont {Irman}}, \bibinfo {author} {\bibfnamefont
  {U.}~\bibnamefont {Schramm}}, \ and\ \bibinfo {author} {\bibfnamefont
  {S.}~\bibnamefont {Karsch}},\ }\href {\doibase 10.1063/1.5041755} {\bibfield
  {journal} {\bibinfo  {journal} {Review of Scientific Instruments}\ }\textbf
  {\bibinfo {volume} {89}},\ \ \bibinfo {pages} {093303} (\bibinfo {year}
  {2018})}\BibitemShut {NoStop}%
\bibitem [{\citenamefont {Decker}\ \emph {et~al.}(1996)\citenamefont {Decker},
  \citenamefont {Mori}, \citenamefont {Tzeng},\ and\ \citenamefont
  {Katsouleas}}]{Decker1996ThePlasmas}%
  \BibitemOpen
  \bibfield  {author} {\bibinfo {author} {\bibfnamefont {C.~D.}\ \bibnamefont
  {Decker}}, \bibinfo {author} {\bibfnamefont {W.~B.}\ \bibnamefont {Mori}},
  \bibinfo {author} {\bibfnamefont {K.~C.}\ \bibnamefont {Tzeng}}, \ and\
  \bibinfo {author} {\bibfnamefont {T.}~\bibnamefont {Katsouleas}},\ }\href
  {\doibase 10.1063/1.872001} {\bibfield  {journal} {\bibinfo  {journal}
  {Physics of Plasmas}\ }\textbf {\bibinfo {volume} {3}},\ \unskip\ \bibinfo
  {pages} {2047--2056} (\bibinfo {year} {1996})}\BibitemShut {NoStop}%
\bibitem [{\citenamefont {Zhu}, \citenamefont {Palastro},\ and\ \citenamefont
  {Antonsen}(2012)}]{Zhu2012StudiesSimulations}%
  \BibitemOpen
  \bibfield  {author} {\bibinfo {author} {\bibfnamefont {W.}~\bibnamefont
  {Zhu}}, \bibinfo {author} {\bibfnamefont {J.~P.}\ \bibnamefont {Palastro}}, \
  and\ \bibinfo {author} {\bibfnamefont {T.~M.}\ \bibnamefont {Antonsen}},\
  }\href {\doibase 10.1063/1.4773722} {\bibfield  {journal} {\bibinfo
  {journal} {Physics of Plasmas}\ }\textbf {\bibinfo {volume} {19}},\ \
  \bibinfo {pages} {033105} (\bibinfo {year} {2012})}\BibitemShut {NoStop}%
\bibitem [{\citenamefont {Raj}\ \emph {et~al.}(2020)\citenamefont {Raj},
  \citenamefont {Kononenko}, \citenamefont {Doche}, \citenamefont {Davoine},
  \citenamefont {Caizergues}, \citenamefont {Chang}, \citenamefont
  {Couperus~Cabada}, \citenamefont {Debus}, \citenamefont {Ding}, \citenamefont
  {F{\"{o}}rster}, \citenamefont {Gilljohann}, \citenamefont {Goddet},
  \citenamefont {Heinemann}, \citenamefont {Kluge}, \citenamefont {Kurz},
  \citenamefont {Pausch}, \citenamefont {Rousseau}, \citenamefont {San
  Miguel~Claveria}, \citenamefont {Sch{\"{o}}bel}, \citenamefont {Siciak},
  \citenamefont {Steiniger}, \citenamefont {Tafzi}, \citenamefont {Yu},
  \citenamefont {Hidding}, \citenamefont {Martinez de~la Ossa}, \citenamefont
  {Irman}, \citenamefont {Karsch}, \citenamefont {D{\"{o}}pp}, \citenamefont
  {Schramm}, \citenamefont {Gremillet},\ and\ \citenamefont
  {Corde}}]{Raj2020ProbingInteractions}%
  \BibitemOpen
  \bibfield  {author} {\bibinfo {author} {\bibfnamefont {G.}~\bibnamefont
  {Raj}}, \bibinfo {author} {\bibfnamefont {O.}~\bibnamefont {Kononenko}},
  \bibinfo {author} {\bibfnamefont {A.}~\bibnamefont {Doche}}, \bibinfo
  {author} {\bibfnamefont {X.}~\bibnamefont {Davoine}}, \bibinfo {author}
  {\bibfnamefont {C.}~\bibnamefont {Caizergues}}, \bibinfo {author}
  {\bibfnamefont {Y.~Y.}\ \bibnamefont {Chang}}, \bibinfo {author}
  {\bibfnamefont {J.~P.}\ \bibnamefont {Couperus~Cabada}}, \bibinfo {author}
  {\bibfnamefont {A.}~\bibnamefont {Debus}}, \bibinfo {author} {\bibfnamefont
  {H.}~\bibnamefont {Ding}}, \bibinfo {author} {\bibfnamefont {M.}~\bibnamefont
  {F{\"{o}}rster}}, \bibinfo {author} {\bibfnamefont {M.~F.}\ \bibnamefont
  {Gilljohann}}, \bibinfo {author} {\bibfnamefont {J.~P.}\ \bibnamefont
  {Goddet}}, \bibinfo {author} {\bibfnamefont {T.}~\bibnamefont {Heinemann}},
  \bibinfo {author} {\bibfnamefont {T.}~\bibnamefont {Kluge}}, \bibinfo
  {author} {\bibfnamefont {T.}~\bibnamefont {Kurz}}, \bibinfo {author}
  {\bibfnamefont {R.}~\bibnamefont {Pausch}}, \bibinfo {author} {\bibfnamefont
  {P.}~\bibnamefont {Rousseau}}, \bibinfo {author} {\bibfnamefont
  {P.}~\bibnamefont {San Miguel~Claveria}}, \bibinfo {author} {\bibfnamefont
  {S.}~\bibnamefont {Sch{\"{o}}bel}}, \bibinfo {author} {\bibfnamefont
  {A.}~\bibnamefont {Siciak}}, \bibinfo {author} {\bibfnamefont
  {K.}~\bibnamefont {Steiniger}}, \bibinfo {author} {\bibfnamefont
  {A.}~\bibnamefont {Tafzi}}, \bibinfo {author} {\bibfnamefont
  {S.}~\bibnamefont {Yu}}, \bibinfo {author} {\bibfnamefont {B.}~\bibnamefont
  {Hidding}}, \bibinfo {author} {\bibfnamefont {A.}~\bibnamefont {Martinez
  de~la Ossa}}, \bibinfo {author} {\bibfnamefont {A.}~\bibnamefont {Irman}},
  \bibinfo {author} {\bibfnamefont {S.}~\bibnamefont {Karsch}}, \bibinfo
  {author} {\bibfnamefont {A.}~\bibnamefont {D{\"{o}}pp}}, \bibinfo {author}
  {\bibfnamefont {U.}~\bibnamefont {Schramm}}, \bibinfo {author} {\bibfnamefont
  {L.}~\bibnamefont {Gremillet}}, \ and\ \bibinfo {author} {\bibfnamefont
  {S.}~\bibnamefont {Corde}},\ }\href {\doibase
  10.1103/physrevresearch.2.023123} {\bibfield  {journal} {\bibinfo  {journal}
  {Phys. Rev. Research}\ }\textbf {\bibinfo {volume} {2}},\ \ \bibinfo {pages}
  {023123} (\bibinfo {year} {2020})}\BibitemShut {NoStop}%
\bibitem [{\citenamefont {Schumaker}\ \emph {et~al.}(2013)\citenamefont
  {Schumaker}, \citenamefont {Nakanii}, \citenamefont {McGuffey}, \citenamefont
  {Zulick}, \citenamefont {Chyvkov}, \citenamefont {Dollar}, \citenamefont
  {Habara}, \citenamefont {Kalintchenko}, \citenamefont {Maksimchuk},
  \citenamefont {Tanaka}, \citenamefont {Thomas}, \citenamefont {Yanovsky},\
  and\ \citenamefont {Krushelnick}}]{Schumaker2013UltrafastInteractions}%
  \BibitemOpen
  \bibfield  {author} {\bibinfo {author} {\bibfnamefont {W.}~\bibnamefont
  {Schumaker}}, \bibinfo {author} {\bibfnamefont {N.}~\bibnamefont {Nakanii}},
  \bibinfo {author} {\bibfnamefont {C.}~\bibnamefont {McGuffey}}, \bibinfo
  {author} {\bibfnamefont {C.}~\bibnamefont {Zulick}}, \bibinfo {author}
  {\bibfnamefont {V.}~\bibnamefont {Chyvkov}}, \bibinfo {author} {\bibfnamefont
  {F.}~\bibnamefont {Dollar}}, \bibinfo {author} {\bibfnamefont
  {H.}~\bibnamefont {Habara}}, \bibinfo {author} {\bibfnamefont
  {G.}~\bibnamefont {Kalintchenko}}, \bibinfo {author} {\bibfnamefont
  {A.}~\bibnamefont {Maksimchuk}}, \bibinfo {author} {\bibfnamefont {K.~A.}\
  \bibnamefont {Tanaka}}, \bibinfo {author} {\bibfnamefont {A.~G.}\
  \bibnamefont {Thomas}}, \bibinfo {author} {\bibfnamefont {V.}~\bibnamefont
  {Yanovsky}}, \ and\ \bibinfo {author} {\bibfnamefont {K.}~\bibnamefont
  {Krushelnick}},\ }\href {\doibase 10.1103/PhysRevLett.110.015003} {\bibfield
  {journal} {\bibinfo  {journal} {Physical Review Letters}\ }\textbf {\bibinfo
  {volume} {110}},\ \ \bibinfo {pages} {015003} (\bibinfo {year}
  {2013})}\BibitemShut {NoStop}%
\bibitem [{\citenamefont {Esarey}, \citenamefont {Ride},\ and\ \citenamefont
  {Sprangle}(1993)}]{Esarey1993NonlinearPlasmas}%
  \BibitemOpen
  \bibfield  {author} {\bibinfo {author} {\bibfnamefont {E.}~\bibnamefont
  {Esarey}}, \bibinfo {author} {\bibfnamefont {S.}~\bibnamefont {Ride}}, \ and\
  \bibinfo {author} {\bibfnamefont {P.}~\bibnamefont {Sprangle}},\ }\href@noop
  {} {\bibfield  {journal} {\bibinfo  {journal} {Physical Review E}\ }\textbf
  {\bibinfo {volume} {48}},\ \unskip\ \bibinfo {pages} {3003--3021} (\bibinfo
  {year} {1993})}\BibitemShut {NoStop}%
\bibitem [{\citenamefont {Tsai}\ \emph {et~al.}(2017)\citenamefont {Tsai},
  \citenamefont {Arefiev}, \citenamefont {Shaw}, \citenamefont {Stark},
  \citenamefont {Wang}, \citenamefont {Zgadzaj},\ and\ \citenamefont
  {Downer}}]{Tsai2017Self-aligningFocus}%
  \BibitemOpen
  \bibfield  {author} {\bibinfo {author} {\bibfnamefont {H.~E.}\ \bibnamefont
  {Tsai}}, \bibinfo {author} {\bibfnamefont {A.~V.}\ \bibnamefont {Arefiev}},
  \bibinfo {author} {\bibfnamefont {J.~M.}\ \bibnamefont {Shaw}}, \bibinfo
  {author} {\bibfnamefont {D.~J.}\ \bibnamefont {Stark}}, \bibinfo {author}
  {\bibfnamefont {X.}~\bibnamefont {Wang}}, \bibinfo {author} {\bibfnamefont
  {R.}~\bibnamefont {Zgadzaj}}, \ and\ \bibinfo {author} {\bibfnamefont
  {M.~C.}\ \bibnamefont {Downer}},\ }\href {\doibase 10.1063/1.4973432}
  {\bibfield  {journal} {\bibinfo  {journal} {Physics of Plasmas}\ }\textbf
  {\bibinfo {volume} {24}},\ \ \bibinfo {pages} {013106} (\bibinfo {year}
  {2017})}\BibitemShut {NoStop}%
\end{thebibliography}%

\end{document}